\documentclass[pdftex, prl, twocolumn, floatfix, superscriptaddress]{revtex4}

\usepackage{times}
\usepackage{amsfonts}
\usepackage{amssymb}
\usepackage{amsmath}
\usepackage{graphicx}
\usepackage{ulem}
\usepackage[colorlinks=true, breaklinks=true, linkcolor=blue, citecolor=blue, urlcolor=blue, pdfborder={0 0 0}]{hyperref}

\newcommand{\doi}[2]{\href{http://dx.doi.org/#1}{#2}}
\newcommand{\arxiv}[1]{\href{http://arxiv.org/abs/#1}{arXiv:#1}}

\begin{document}

\title{Antiferromagnetic Heisenberg Spin Chain of a Few Cold Atoms in a One-Dimensional Trap}

\author{S.~Murmann}
\email{murmann@physi.uni-heidelberg.de}
\affiliation{Physikalisches Institut der Universit\"at Heidelberg, Im Neuenheimer Feld 226, DE-69120 Heidelberg, Germany}

\author{F.~Deuretzbacher}
\email{frank.deuretzbacher@itp.uni-hannover.de}
\affiliation{Institut f\"ur Theoretische Physik, Leibniz Universit\"at Hannover, Appelstra{\ss}e 2, DE-30167 Hannover, Germany}

\author{G.~Z\"urn}
\affiliation{Physikalisches Institut der Universit\"at Heidelberg, Im Neuenheimer Feld 226, DE-69120 Heidelberg, Germany}

\author{J.~Bjerlin}
\affiliation{Mathematical Physics and NanoLund, LTH, Lund University, SE-22100 Lund, Sweden}

\author{S.~M.~Reimann}
\affiliation{Mathematical Physics and NanoLund, LTH, Lund University, SE-22100 Lund, Sweden}

\author{L.~Santos}
\affiliation{Institut f\"ur Theoretische Physik, Leibniz Universit\"at Hannover, Appelstra{\ss}e 2, DE-30167 Hannover, Germany}

\author{T.~Lompe}
\thanks{Present address: Institut f\"ur Laserphysik, Universit\"at Hamburg, Luruper Chaussee 149, DE-22761 Hamburg, Germany}
\affiliation{Physikalisches Institut der Universit\"at Heidelberg, Im Neuenheimer Feld 226, DE-69120 Heidelberg, Germany}

\author{S.~Jochim}
\affiliation{Physikalisches Institut der Universit\"at Heidelberg, Im Neuenheimer Feld 226, DE-69120 Heidelberg, Germany}

\begin{abstract}

We report on the deterministic preparation of antiferromagnetic Heisenberg spin chains consisting of up to four fermionic atoms in a one-dimensional trap. These chains are stabilized by strong repulsive interactions between the two spin components without the need for an external periodic potential. We independently characterize the spin configuration of the chains by measuring the spin orientation of the outermost particle in the trap and by projecting the spatial wave function of one spin component on single-particle trap levels. Our results are in good agreement with a spin-chain model for fermionized particles and with numerically exact diagonalizations of the full few-fermion system.

\end{abstract}

\maketitle

The high control and tunability of ultracold atomic systems offer the fascinating possibility to simulate quantum magnetism~\cite{Lewenstein07}, a topic of fundamental importance in condensed matter physics~\cite{Auerbach}. Systems of spin-1/2 fermions with antiferromagnetic (AFM) correlations are of particular interest due to the observation of high-temperature superconductivity in cuprates with AFM correlations~\cite{Norman11}. The experimental implementation of the necessary exchange couplings is usually realized by superexchange processes of neighboring atoms in the Mott-insulating state of a deep optical lattice. Superexchange couplings were measured in both bosonic~\cite{Trotzky08} and fermionic double-well systems~\cite{Murmann15} and short-range AFM correlations of fermionic atoms were detected in various lattice geometries~\cite{Greif13, Hart15, Messer15}. Furthermore, superexchange processes were used to study the dynamics of spin impurities above the ferromagnetic (FM) ground state of bosons in the Mott-insulating state of a one-dimensional (1D) lattice~\cite{Fukuhara13a}. Bosonic atoms were also used to simulate AFM Ising spin chains in a tilted optical lattice~\cite{Simon11, Meinert13}. However, the AFM ground state of spin-1/2 fermions in a deep optical lattice has so far not been realized due to the very low energy scale associated with the superexchange coupling.

\begin{figure}[b!]
\begin{center}
\includegraphics[width = \columnwidth]{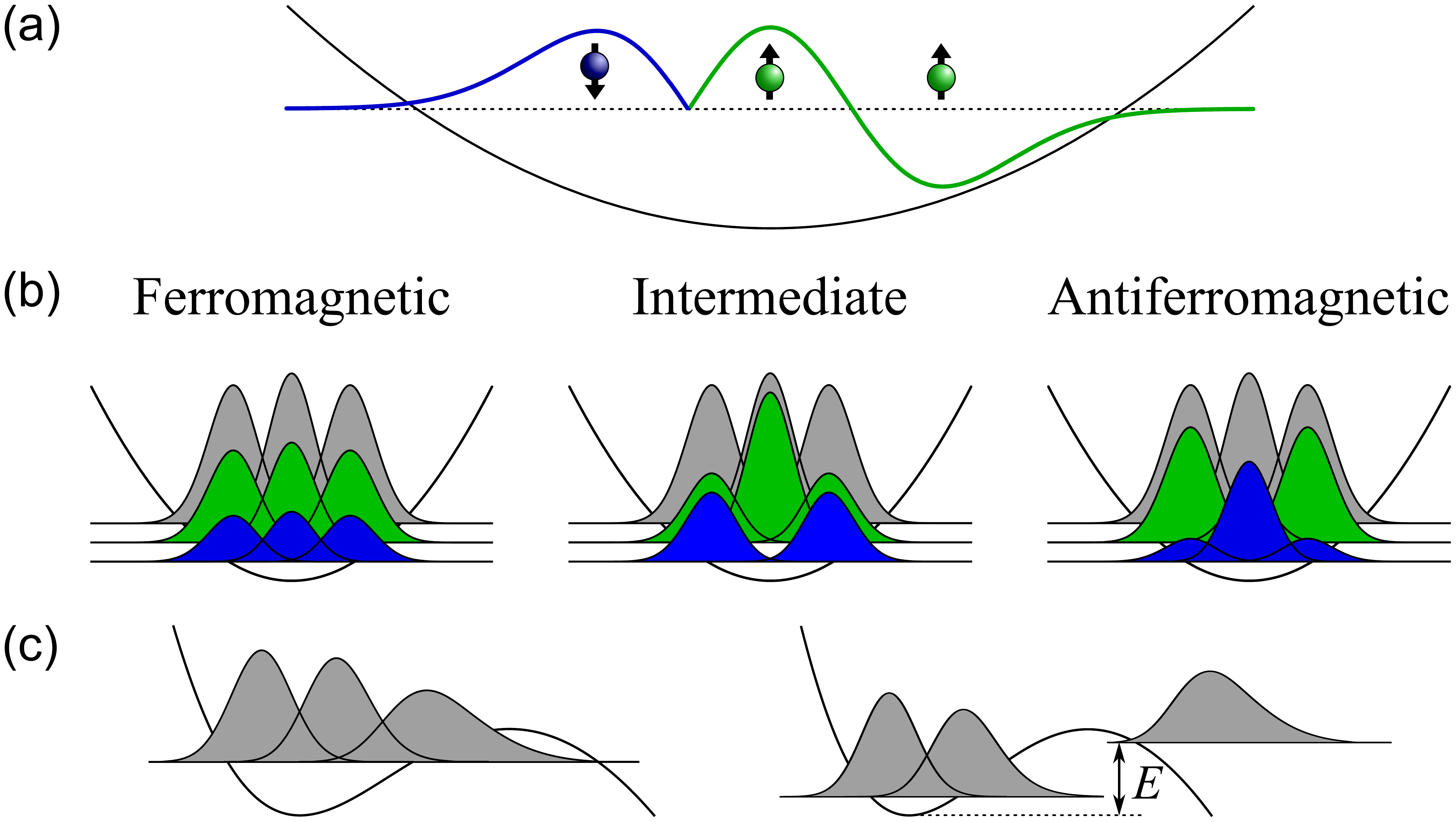}
\caption{Heisenberg spin chain of three fermions. (a) Sketch of two spin-up and one spin-down atom with diverging 1D coupling constant ($g_\text{1D} = \pm \infty$) in a harmonic trap. If the relative spatial wave function of two distinguishable fermions is symmetric, the strong interactions induce a cusp in the relative wave function of the two particles (left side). This causes them to separate like identical fermions (right side). In this fermionization limit the system forms a Wigner-crystal-like state with fixed ordering of the particles. (b) Single-particle contributions to the total (gray), the spin-up (green), and the spin-down density (blue) of two spin-up and one spin-down atom in the fermionization regime in a harmonic trap. Like in a Wigner crystal, the total densities of the ferromagnetic (left), the intermediate (middle), and the antiferromagnetic state (right) are identical, while their spin densities differ and are determined by a Heisenberg spin-chain Hamiltonian. (c) Densities of three particles before (left) and after (right) the tunneling of one atom with energy $E$ out of a tilted trap. At fermionization, only the rightmost particle can leave the trap in the tunneling process.}
\label{fig-sketch}
\end{center}
\end{figure}

This problem can be circumvented in 1D systems, where quantum magnetism can be simulated without an optical lattice~\cite{Deuretzbacher14, Beverland14, Pagano14}. In the regime of strong interactions, the spatial wave function of both fermions~\cite{Zuern12} and bosons~\cite{Girardeau60, Paredes04, Kinoshita04} can be mapped on the wave function of spinless noninteracting fermions [Fig.~\ref{fig-sketch}(a)]. In this so called fermionization limit, the strong interactions lead to the formation of a Wigner-crystal-like state~\cite{Matveev04, Deuretzbacher08, Matveev08}, which has a highly degenerate ground state when the particles have multiple internal degrees of freedom [Fig.~\ref{fig-sketch}(b)]~\cite{Girardeau07, Deuretzbacher08, Matveev08, Guan09}. Close to the limit of fermionization, the structure of the quasi-degenerate ground-state multiplet~\cite{Bugnion13, Gharashi13, Sowinski13, Cui13, Lindgren13, Volosniev14, Harshman14, Garcia14, Yang15, Levinsen15} is determined by an effective Sutherland spin-chain Hamiltonian, which for two-component systems becomes a Heisenberg model~\cite{Deuretzbacher14, Matveev04, Matveev08, Guan08, Volosniev14, Yang15, Levinsen15}.

In this paper, we report on the realization of Heisenberg spin chains of $N_\uparrow$ spin-up and $N_\downarrow$ spin-down particles with $(N_\uparrow, N_\downarrow) = (2,1)$, ${(3,1)}$, and ${(2,2)}$.
% We show that the noninteracting ground states of these systems are adiabatically connected to the respective AFM states of the effective spin chain Hamiltonians in the fermionization regime~\cite{Sowinski13, Gharashi13, Lindgren13, Harshman14}.
We show that under an adiabatic change of the interaction strength the noninteracting ground states of these systems evolve into the respective AFM states in the limit of infinitely strong repulsion~\cite{Gharashi13, Lindgren13}.
We identify the AFM states by two independent measurements. First, we use a tunneling technique to measure the spin orientation of the outermost particle of the spin chain. Second, we probe the spatial wave function of the spin-down atom in the $(2,1)$ and $(3,1)$ system by projecting it on single-particle trap levels.

In our experiments, we realize a spin-$1/2$ system by trapping ultracold $^6\text{Li}$ atoms in an elongated optical dipole trap~\cite{Supplements, Serwane11} in their two lowest hyperfine states ${\left|\uparrow\right\rangle \equiv \left|j=1/2, m_j=-1/2; I=1, m_I=0\right \rangle}$ and ${\left|\downarrow\right\rangle \equiv \left|j=1/2, m_j=-1/2; I=1, m_I=1\right \rangle}$. As the energy of the atoms is much smaller than the lowest transverse excitation energy in the trap, their dynamics are restricted to the longitudinal axis of the trap. In such a quasi-1D system, the interaction strength between ultracold atoms of opposite spin is determined by the 1D coupling constant, $g_\text{1D}$, which diverges at a confinement-induced resonance (CIR) when the 3D scattering length, $a_\text{3D}$, approaches the harmonic oscillator length of the radial confinement~\cite{Supplements, Olshanii98}. We use a magnetic Feshbach resonance to control $a_\text{3D}$ and therefore are able to smoothly tune $g_\text{1D}$ across the CIR. At the same time, scattering between fermionic atoms of the same spin component is forbidden. Throughout this paper, $g_\text{1D}$ will be given in units of $a_{||}\hbar\omega_{||}$, where $a_{||} = \sqrt{\hbar / m \omega_{||}}$ and $\omega_{||}$ are the harmonic oscillator length and the trap frequency in longitudinal direction and $m$ is the mass of a $^6\text{Li}$ atom. 

\begin{figure}[ht!]
\begin{center}
\includegraphics[width = \columnwidth]{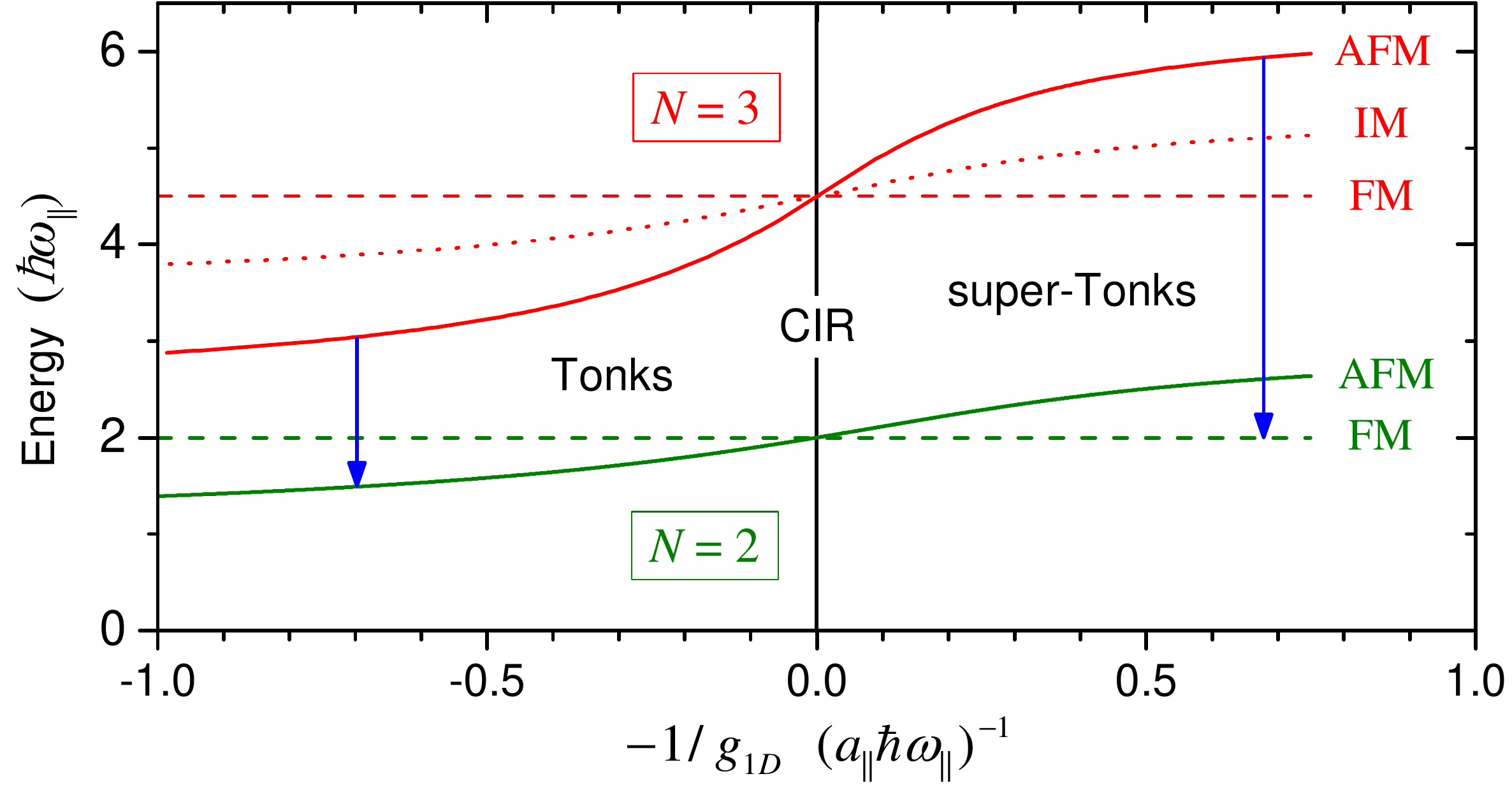}
\caption{Energies in spin-chain regime. Eigenenergies of two (green) and three (red) strongly interacting spin-1/2 fermions in a 1D harmonic trap as a function of the interaction strength. In the Tonks regime, the antiferromagnetic states are the ground states of each multiplet, while the ferromagnetic states have the highest energies. In the super-Tonks regime, the ordering of the energy levels is inverted. Close to the confinement-induced resonance (CIR), the energy shifts are linear in $-1/g_\text{1D}$ and can be determined by a Heisenberg spin-chain Hamiltonian. The system is initially prepared in the noninteracting ground state of the three-particle system at $-1/g_\text{1D} = -\infty$,
% which is adiabatically connected to the antiferromagnetic state (red solid line).
which evolves, for increasing $-1/g_\text{1D}$, into the antiferromagnetic state around the CIR (red solid line).
During a ramp across the CIR, the system stays in the antiferromagnetic state, since all eigenstates of the system are decoupled. The blue arrows indicate the predominant channels for the tunneling of one atom below (left) and above (right) the fermionization regime.}
\label{fig-tunneling_energies}
\end{center}
\end{figure}

We start our experiments by preparing a $(2,1)$, $(3,1)$, or $(2,2)$ system in the noninteracting many-particle ground state of the trap~\cite{Supplements, Serwane11}. By changing the magnetic offset field with a constant rate, we ramp the system into the fermionization regime close to the CIR (Fig.~\ref{fig-tunneling_energies}), where it forms a spin chain. Below the CIR, $g_\text{1D}$ is large and positive and the system is in the Tonks regime of strong repulsion~\cite{Paredes04, Kinoshita04}. When crossing the CIR, $g_\text{1D}$ changes sign from $+\infty$ to $-\infty$ while the system continuously follows the so called upper branch~\cite{Gharashi13} into the super-Tonks regime of strong attraction~\cite{Astrakharchik05, Haller09, Zuern12} (Fig.~\ref{fig-tunneling_energies}). In the super-Tonks regime, the system is in an excited state, which is metastable against decay into bound states.

\begin{figure}[ht!]
\begin{center}
\includegraphics[width = \columnwidth]{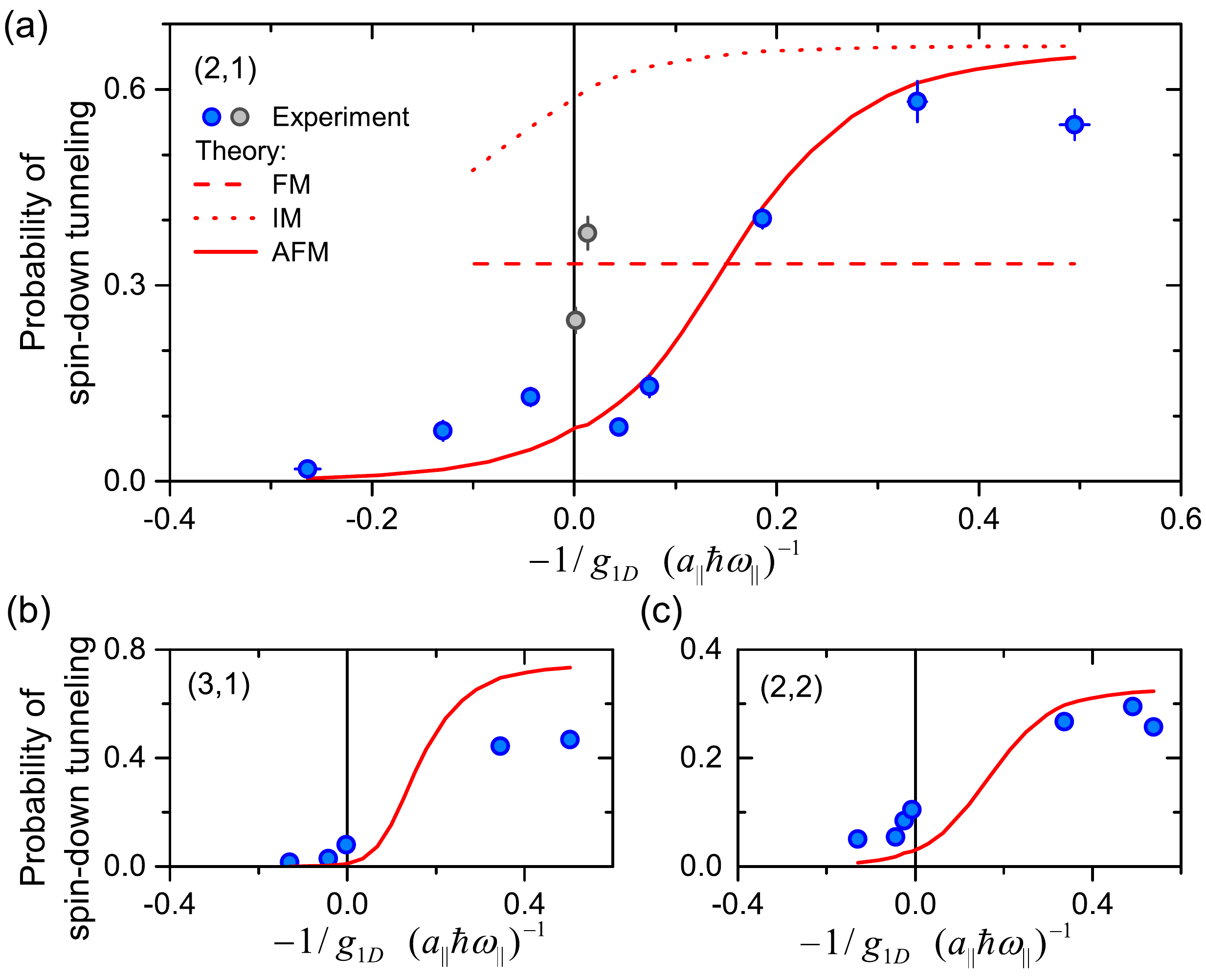}
\caption{Probing the spin distribution. Tunneling probabilities of the spin-down atom in a $(2,1)$ system (a) and a $(3,1)$ system (b) and tunneling probability of both spin-down atoms in a $(2,2)$ system (c) as a function of the interaction strength. The red lines are the solutions of a tunneling model for the antiferromagnetic (solid), the ferromagnetic (dashed), and the intermediate state (dotted).  The gray points in (a) indicate a narrow resonance between the antiferromagnetic and the intermediate state of the $(2,1)$ system close to $-1/g_\text{1D}=0$.
% \textcolor{red}{The red crosses in (b) show the calculated probability of spin-down tunneling, which was corrected for the fact that the energy of the tunneling atom in these measurements exceeded the height of the tunneling barrier.}
Error bars denote the $1 \sigma$ statistical uncertainties.}
\label{fig-tunneling}
\end{center}
\end{figure}

In a first set of measurements, we identify the states of the spin chains by probing the spin distributions in the trap. Here, we make use of the fact that in the fermionization regime the atoms become impenetrable and therefore their ordering along the longitudinal axis of the trap is fixed. This allows us to determine the spin orientation of the outermost particle in the trap in a tunneling measurement. To do this, we tilt the trap as shown in Fig.~\ref{fig-sketch}(c) and thereby allow atoms to tunnel out of the trap. We carefully adjust the trap parameters during the tunneling process, to let exactly one atom (for the $(2,1)$ and the $(3,1)$ systems) or two atoms (for the $(2,2)$ system) tunnel~\cite{Supplements}. Finally, we measure the number of spin-up atoms in the final state to determine the spin of the atoms that left the trap during the tunneling process~\cite{Supplements}. We define spin-down tunneling as the process in which all spin-down atoms tunnel out. By repeating this measurement at different magnetic offset fields, we deduce the probability of spin-down tunneling, $P_\downarrow(-1/g_\text{1D})$, as a function of the inverse 1D coupling constant as shown in Fig.~\ref{fig-tunneling}.

As shown in Fig.~\ref{fig-sketch}(b) for a $(2,1)$ system in a harmonic trap, the different states of the spin chain can be uniquely identified by their spin densities~\cite{Deuretzbacher14}, specifically by the probability of the outermost spin to point downwards. Since in the fermionization regime the ordering of the atoms in the trap is fixed only the outermost atom can escape during the tunneling process. Exactly at the CIR, the probability of spin-down tunneling should therefore directly reveal the state of the spin chain~\cite{Volosniev14, Deuretzbacher14, Levinsen15}. Away from resonance, the probability of spin-down tunneling is also influenced by the energy of the final in-trap states, favoring final states with lower energy as indicated by the blue arrows in Fig.~\ref{fig-tunneling_energies}. To identify the spin states throughout the entire spin-chain regime, we compare our data to the results of a tunneling model, which in the following section is explained for a $(2,1)$ system.

In our tunneling model, the initial states are eigenstates of a Heisenberg spin-chain Hamiltonian~\cite{Supplements}, where the exchange couplings $J_i$ between neighboring spins depend on the trap geometry and on the inverse 1D coupling constant~\cite{Deuretzbacher14}. For the $(2,1)$ system with repulsive interactions and a symmetric trap $(J_1=J_2>0)$, these eigenstates are the antiferromagnetic (AFM) ground state, the intermediate (IM) state, and the ferromagnetic (FM) state as shown in Fig.~\ref{fig-sketch}(b). During the tunneling process the trap is tilted as shown in Fig.~\ref{fig-sketch}(c) and therefore the density is not symmetric. Hence, the exchange couplings are not identical anymore $(|J_1|>|J_2|)$, which leads to a coherent mixing of the AFM and IM state during the tunneling process~\cite{Supplements}. We calculate a probability of approximately $8\%$ for the rightmost spin of the AFM state in the tilted trap to point downwards. This is in good agreement with the blue data points in Fig.~\ref{fig-tunneling}(a) that cross the CIR at $P_\downarrow \approx 10\%$.

Away from the CIR, the eigenstates of both the three-particle and the two-particle spin chains are nondegenerate (Fig.~\ref{fig-tunneling_energies}). In this case, the energies of the initial three-particle state $\left|i\right\rangle$ and the final two-particle state $\left|f\right\rangle$ involved in the tunneling process are important, since their difference determines the energy $E$ of the tunneling particle. The tunneling rate of the particle that leaves the trap is strongly affected by its energy and can be calculated as
\begin{equation} \label{eq-tunnel_rate}
T_{i,f} \propto |\langle i|f,t\rangle|^2 E e^{-2\gamma(E)} ,
\end{equation}
where $\left|f,t\right\rangle = \left|f\right\rangle \otimes \left|t\right\rangle$ with $\left|t\right\rangle$ indicating the spin orientation of the tunneling particle. The tunneling parameter $\gamma$ is determined by means of a WKB calculation~\cite{Supplements}. The probability to tunnel from state $\left|i\right\rangle$ to state $\left|f\right\rangle$ is given by
\begin{equation} \label{eq-tunnel_probability}
P_{i,f} = \frac{T_{i,f}}{( \sum_{f'} T_{i,f'} )} ,
\end{equation}
where the sum is over all possible final states $\left|f'\right\rangle$.

Using Eq.~(\ref{eq-tunnel_probability}), we calculate the probabilities $P_{i,|\uparrow,\uparrow\rangle}$ of tunneling into the spin-polarized final state [red lines in Fig.~\ref{fig-tunneling}(a)], which is equivalent to the probability of spin-down tunneling ($P_\downarrow$). Far below the CIR, the energy dependent term $E e^{-2\gamma}$ dominates the outcome of the tunneling rates (Eq.~\ref{eq-tunnel_rate}). Therefore, tunneling into the AFM two-particle ground state $({\left|\uparrow,\downarrow\right\rangle}-{\left|\downarrow,\uparrow\right\rangle})/\sqrt{2}$ is strongly favored if its spin overlap to the initial state is not zero. This leads to a limiting value of $P_\downarrow = 0$ for initial AFM and IM states. Above the resonance, the energy ordering of the two-particle FM and AFM states is reversed and tunneling into the FM states is predominant (Fig.~\ref{fig-tunneling_energies})~\footnote{At the values of $g_\text{1D}$ attainable in our experiment, the spatial wave function overlap between the spin-chain states and the molecular ground state is negligible. We expect that for a small negative $g_\text{1D}$ tunneling into the molecular state is predominant.}. Here, $P_\downarrow$ is determined by the ratio of the spin overlaps between the first two spins of the initial states and the FM two-particle states ${\left|\uparrow,\uparrow\right\rangle}$ and $({\left|\uparrow,\downarrow\right\rangle}+{\left|\downarrow,\uparrow\right\rangle})/\sqrt{2}$.

The comparison of the theoretically predicted $P_\downarrow$ of the AFM state in the tilted trap [red solid line in Fig~\ref{fig-tunneling}(a)] with the experimental data (blue points) shows good agreement, while the FM (red dashed line) and IM (red dotted line) states are clearly excluded. We therefore conclude that before tunneling both below and above the CIR the system is in the AFM state. The gray points at $-1/g_\text{1D} \approx 0$ indicate a narrow resonance effect that couples the AFM state to the IM state of the spin chain. Since this resonance is accompanied by strongly enhanced three-body losses~\cite{Supplements}, we suspect it to be caused by a coupling of the AFM and the IM states via a molecular state with center-of-mass excitation. The coupling to such molecular states is strongly enhanced by the anharmonicity of our tilted trap~\cite{Sala13}.

For the AFM state of the $(3,1)$ system, a similar calculation predicts $P_\downarrow \approx 1\%$ on resonance and a saturation value of $P_\downarrow \approx 75\%$ deep in the super-Tonks regime~\cite{Supplements}. As shown in Fig~\ref{fig-tunneling}(b), the general trend of our measurements agrees with this prediction for the AFM state, but in the super-Tonks regime, there is a significant deviation.
%\textcolor{blue}{The reason for this deviation is that the calculation assumes an adiabatic lowering of the potential barrier. As a result, the tunneling energies of all tunneling channels are always well below the barrier maximum. We believe that this condition is not fulfilled for the $(3,1)$ system in the super-Tonks regime. Indeed, if we model a nonadiabatic lowering of the potential barrier, the tunneling energy for tunneling into the IM final in-trap state is only slightly below the barrier maximum. As a result, the contribution from the IM tunneling channel leads to a significant reduction of $P_\downarrow$ (38\% and 61\% for the two data points in the super-Tonks regime)~\cite{Supplements}.}
The reason for this deviation is that the calculation assumes an adiabatic lowering of the potential barrier. As a result, the tunneling energies of all tunneling channels are always well below the barrier maximum. We believe that this condition is not fulfilled for the $(3,1)$ system in the super-Tonks regime, where an especially low potential barrier was used for the tunneling measurement. Indeed, if we model a nonadiabatic lowering of the potential barrier, the contribution from tunneling into the IM state reduces $P_\downarrow$ to values that are compatible with the experimental results~\cite{Supplements}.
In order to study the spin configuration of the balanced $(2,2)$ system, we adapt the previous procedure and let two atoms tunnel out of the trap. Here, $P_{\downarrow}$ is defined as the probability to end up in state ${\left|\uparrow,\uparrow\right\rangle}$, where both spin-down atoms tunneled out of the trap. Again, the predicted $P_{\downarrow} \approx 4\%$ on resonance and the limiting value of $P_{\downarrow} \approx 33.3\%$ in the super-Tonks regime are in good agreement with the experiment as shown in Fig.~\ref{fig-tunneling}(c).

To independently confirm the results of our measurement of the spin distribution, we perform a second set of measurements that directly probes the spatial wave function of the system. As shown in Fig.~\ref{fig-sketch}(a), the relative spatial wave function between identical spins always exhibits a smooth zero crossing, while between distinguishable spins with strong interactions it can exhibit a cusp. The cusps lead to occupancies of high-energy trap levels, while the zero crossings require only the occupation of the lowest trap levels. In general, the more symmetric the spatial wave function of a state is, the more cusps it will contain. Therefore, the occupation-number distribution on single-particle trap levels directly reveals the spin configuration of the system.

\begin{figure}[t]
\begin{center}
\includegraphics[width = \columnwidth]{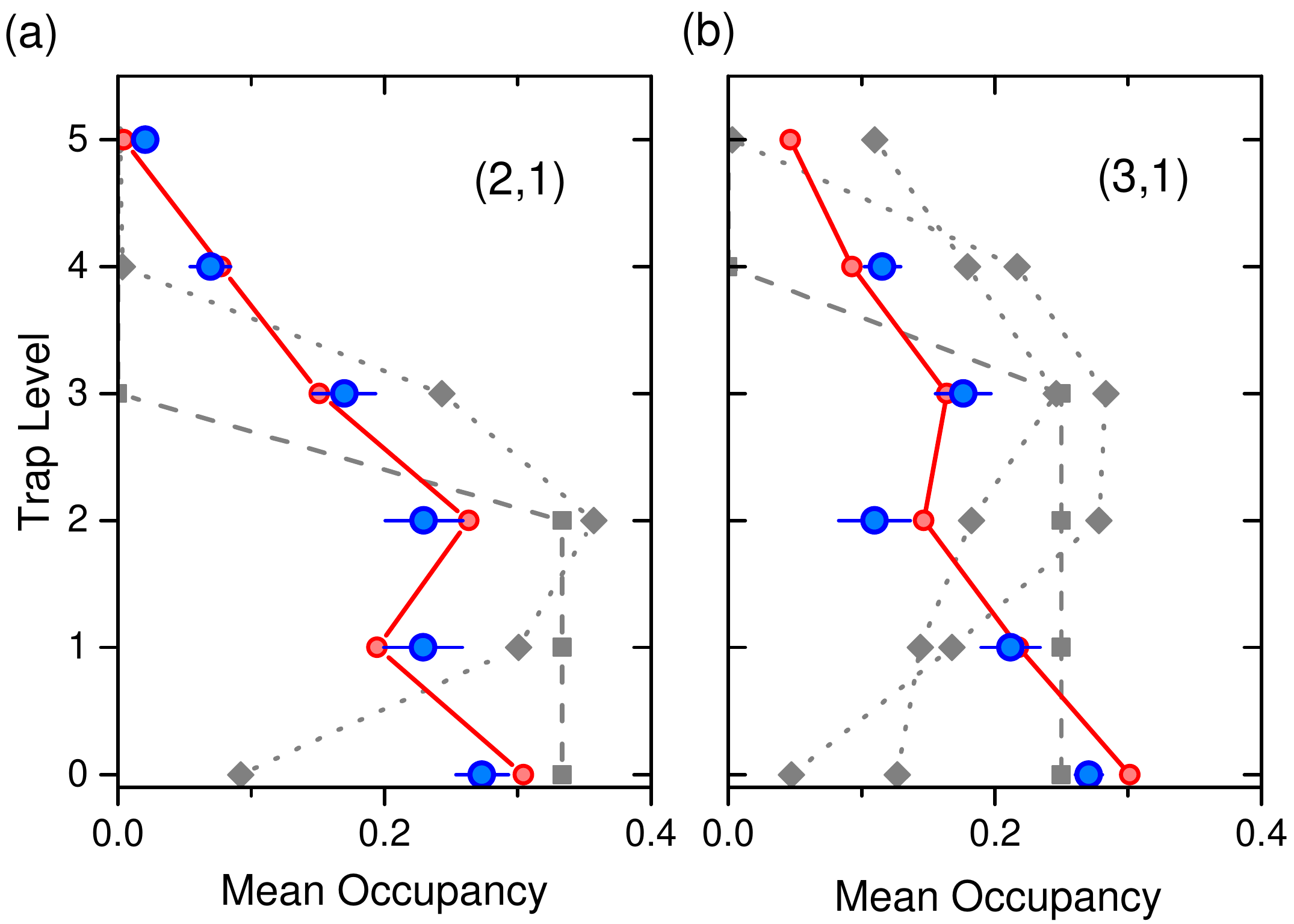}
\caption{Probing the spatial wave function. Occupation-number distribution of the spin-down atom on single-particle trap levels for an initial (a) $(2,1)$ and (b) $(3,1)$ system. The red and gray symbols show theoretical predictions for the antiferromagnetic state (red circles), the ferromagnetic state (gray squares), and intermediate states (gray diamonds). Both measurements (blue points) were made in the super-Tonks regime [$-1/g_\text{1D} = 0.586 \pm 0.014$ for the $(2,1)$ system and $-1/g_\text{1D} = 0.536 \pm 0.013$ for the $(3,1)$ system] to show that both systems stay in the respective antiferromagnetic states throughout the regime of fermionization. Error bars denote the $1 \sigma$ statistical uncertainty.}
\label{fig-occupancies}
\end{center}
\end{figure}

% \sout{To probe this distribution, we prepare an interacting $(2,1)$ or $(3,1)$ system and remove all atoms of the spin-up component from the trap with a $15 \mu\text{s}$ pulse of resonant light.}
To probe this distribution, we prepare an interacting (2,1) or (3,1) system and remove all atoms of the spin-up component from the trap with a short pulse of light. The light is $\sigma^-$ polarized and resonant to the D2 transition of the spin-up atoms $( \left|\uparrow\right\rangle = \left|j=1/2, m_j=-1/2; I=1, m_I=0\right \rangle $ to $\left|j=3/2,m_j=-3/2; I=1, m_I=0\right \rangle) $. We confirm that within our experimental fidelity all spin-up atoms are removed from the trap by the light pulse, while only $3~\%$ of the population of spin-down atoms is lost. With $15 \mu\text{s}$ the duration of the light pulse is significantly shorter than the inverse longitudinal trap frequency of approximately $100 \mu\text{s}$, which sets the timescale of redistribution along the spin chain. This process therefore projects the spin-down component of the wave function of the interacting (${N_\uparrow,1}$)-particle system on single-particle trap levels. Finally, we measure the mean occupancies on the single-particle trap levels~\cite{Supplements}. In Fig.~\ref{fig-occupancies} we compare the mean occupancies of the spin-down atom for the $(2,1)$ and the $(3,1)$ systems in the super-Tonks regime with the theoretical prediction that we obtained by numerically diagonalizing the many-body Hamiltonian for these systems. The comparison shows that both systems are in the AFM spin state and thereby confirms that our systems follow this state throughout the fermionization regime.

In conclusion, we have prepared antiferromagnetic Heisenberg spin chains of up to four atoms in a one-dimensional trap and independently probed the spin distributions and spatial wave functions of the systems. This constitutes a direct observation of quantum magnetism beyond two-particle correlations in a system of ultracold fermionic atoms. By using the methods developed in Ref.~[\onlinecite{Murmann15}], multiple spin chains can be realized and coupled, which offers a new approach to studying two and three-dimensional quantum magnetism.\\ \\

\begin{acknowledgements}

We thank D.~Blume, I.~Brouzos, G.~M.~Bruun, G.~Conduit, J.~C.~Cremon, S.~E.~Gharashi, C.~Greene, M.~Rontani, A.~N.~Wenz and N.~T.~Zinner for helpful discussions and input. This work was supported by the European Research Council starting Grant No.~279697, the Heidelberg Center for Quantum Dynamics, the DFG (project SA 1031/7-1 and RTG 1729), the Cluster of Excellence QUEST, the Swedish Research Council, and NanoLund.

S.M. and F.D. contributed equally to this work.

\end{acknowledgements}

\bibliographystyle{prsty}

\clearpage

\onecolumngrid
\begin{appendix}

\renewcommand\thesection{\arabic{section}}
\renewcommand\thefigure{S\arabic{figure}}
\renewcommand\theequation{S\arabic{equation}}
\setcounter{figure}{0}
\setcounter{equation}{0}

\section{Supplemental material}

\section{I. \quad	Trapping potential}
\label{sec:trap_pot}

\noindent The optical dipole trap (ODT) is created by the focus of a single far red-detuned Gaussian laser beam. For a given laser power $P$, the potential along the propagation direction of the light, which we denote the longitudinal axis, can be written as:
\begin{equation}
V_\text{optical}(z) = p V_0~ \bigg( 1-\frac{1}{1+(z/z_R)^2} \bigg) .
\label{dipole_trap}
\end{equation}
Here, $V_0 = k_\text{B}~3.326~\mu\text{K}$ is the initial depth of the optical potential at a laser power of $P_0 = (265\pm27)~\mu\text{W}$, $p=P/P_0$ is the trap depth parameter and $z_R=\frac{\pi w_0^2}{\lambda}$ is the Rayleigh range of the optical trapping beam with minimal waist $w_0 = 1.838~\mu\text{m}$ and wavelength $\lambda=1064~\text{nm}$. By changing the laser power $P$, we can adjust the trap depth during our experiments. In a harmonic approximation the trap frequencies of the dipole trap are $\omega_{||}=2 \pi \sqrt{p} ~ (1.488 \pm 0.014)~\text{kHz}$ along the longitudinal direction and $\omega_{\perp}=2 \pi \sqrt{p} ~ (14.22 \pm 0.35)~\text{kHz}$ in the radial direction, resulting in an aspect ratio of $\eta \approx 10$.

To allow atoms to tunnel out of the trap, we create a potential barrier by superimposing the optical dipole trap with a linear magnetic gradient along the longitudinal axis [See Fig.~1(c) of main text]. This adds the magnetic part
\begin{equation}
V_\text{magnetic}(z) = -\mu_m B' z
\label{magnetic_gradient}
\end{equation}
to the potential, where $\mu_m$ is the magnetic moment of the atoms and $B' = 18.92~\text{G/cm}$ is the strength of the magnetic field gradient. For detailed information on the determination of the potential parameters, see Ref.~[\onlinecite{Zuern12b}].

\section{II. \quad	Spilling of atoms from the trap}
\label{sec:spilling}

\begin{figure}[t]
\begin{center}
\includegraphics[width = 0.8\columnwidth]{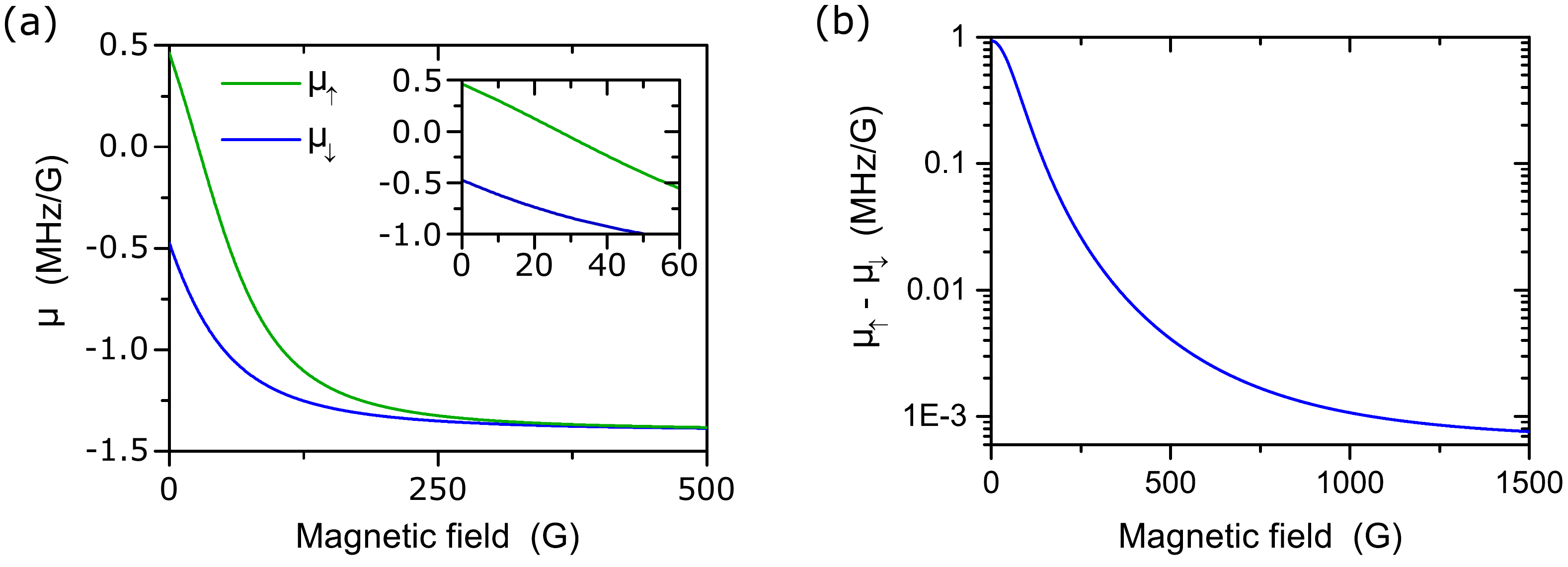}
\caption{Magnetic moment of spin-up and spin-down atoms (a) and difference of the magnetic moments of spin-up and spin-down atoms as a function of the magnetic field.}
\label{fig-mu}
\end{center}
\end{figure}

\noindent In several parts of our experimental sequence, we tilt the potential of the optical dipole trap (ODT) to spill atoms from the trap [Fig.~1(c) of the main text]. To do this, we apply a magnetic field gradient of $B' = 18.92~\text{G/cm}$, which adds a linear potential along the longitudinal axis of the ODT (Section~I). The strength of this linear magnetic potential depends on the magnetic moment of the atoms in the trap, which is shown in Figure~\ref{fig-mu}(a) as a function of the magnetic offset field. By tilting the ODT, we create a potential barrier that separates the in-trap states from the continuum of states outside of the trap. We reduce the height of the potential barrier by lowering the optical power in the ODT ($p<0$). Atoms now leave the trap with rates that strongly depend on their energy. This allows us to remove atoms with energies above or marginally below the potential barrier from the trap while keeping atoms with significantly lower energies. After a certain time we ramp the optical power in the ODT back up to its initial value and thereby stop the spilling process.

When performing this spilling technique in the Paschen-Back regime, the magnetic moments of spin-up and spin-down atoms are approximately equal [Fig.~\ref{fig-mu}(b)]. Therefore, the potential has the same shape for the spin-up and spin-down atoms. All tunneling measurements of the main text are performed deep in the Paschen-Back regime ($\geq 725~\text{G}$), where the magnetic moments of spin-up and spin-down atoms differ by less than $0.15~\%$. We use the spilling technique to let exactly one atom (for the (2,1) and the (3,1) system) or two atoms (for the (2,2) system) tunnel out of the trap (Section~VII).

\section{III. \quad	Preparation of initial states and measurement of atom numbers}
\label{sec:preparation}

\subsection{Initial state preparation}
\noindent To prepare the initial few-particle systems, we start with a degenerate gas of about 600 noninteracting $^6\text{Li}$ atoms in their two lowest hyperfine states in the ODT. At a magnetic offset field of $527~\text{G}$, where the atomic sample is noninteracting and the magnetic moments of spin-up and spin-down atoms are approximately equal, we spill all atoms above a certain trap level from the trap (Section~II). This allows us to prepare spin-balanced ground-state systems of up to 10 atoms with a fidelity of $\sim 90~\%$~[\onlinecite{Serwane11}].

To prepare spin-imbalanced systems, we first prepare a spin-balanced system and then perform a second spilling process at a magnetic offset field of $40~\text{G}$. At this field, the magnetic moment of the spin-up atoms (${F=1/2, m_F=-1/2}$) is negligible [See inset of figure~\ref{fig-mu}(a)], which allows us to selectively spill spin-down atoms (${F=1/2, m_F=+1/2}$) from the trap~[\onlinecite{Serwane11}].

\subsection{Atom number measurement}
\noindent In a single experimental run, we can either measure the total atom number $N$ or the number of spin-up atoms $N_\uparrow$ in the ODT. To measure $N$, we release the atoms from the ODT, recapture them in a magneto-optical trap and measure their fluorescence signal. To measure $N_\uparrow$, we selectively spill all spin-down atoms from the trap and then measure the number of remaining atoms.

A more detailed description of the preparation of both balanced and imbalanced few-fermion systems and of the detection of atom numbers in our experiment can be found in Ref.~[\onlinecite{Serwane11}].

\section{IV. \quad	Calculation of the 1D coupling constant}

\noindent In a gas of ultracold $^6\text{Li}$ atoms, the interactions between the atoms are well described by contact s-wave scattering. In a one-dimensional system, the interaction potential between two atoms in different hyperfine states can be written as:
\begin{equation}
V_\text{int}(z_1-z_2)=g_\text{1D} \, \delta(z_1-z_2),
\end{equation}
where $z_1$ and $z_2$ are the positions of the two atoms, $g_\text{1D}$ is the 1D coupling constant, and $\delta$ is the delta function~[\onlinecite{Olshanii98}]. Contact interactions between identical fermions are forbidden.

\begin{figure}[t]
\begin{center}
\includegraphics[width = 0.5\columnwidth]{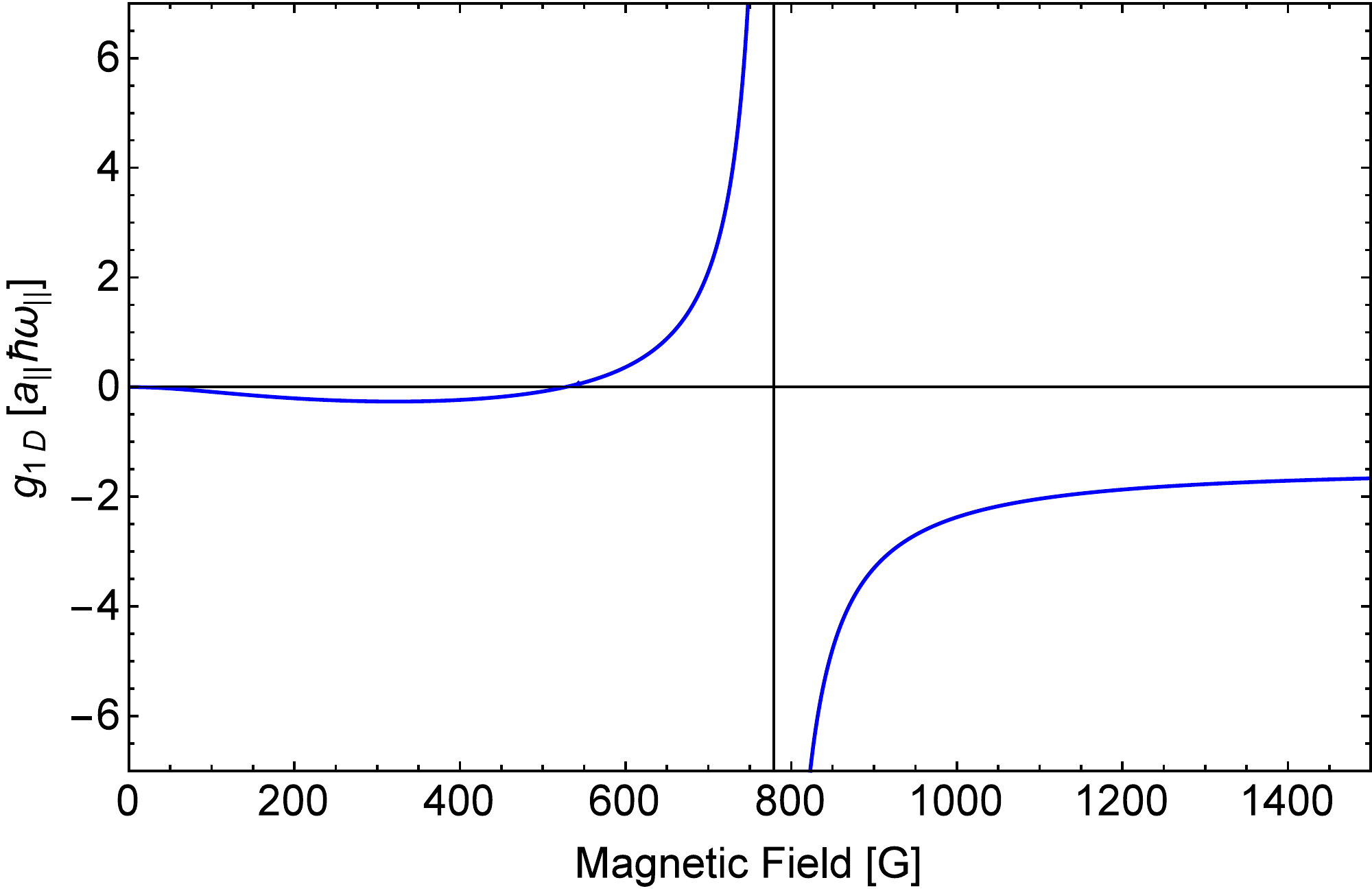}
\caption{Interaction strength of a quasi-one-dimensional system. 1D coupling constant $g_\text{1D}$ between atoms of the two lowest hyperfine states of $^6\text{Li}$ as a function of the offset magnetic field.  At a magnetic field of approximately $779~\text{G}$, the 3D scattering length $a_\text{3D}$ equals $\sqrt{2} a_\perp / C$ and $g_\text{1D}$ diverges in a confinement-induced resonance (Eq.~\ref{eq-gOneD})}
\label{fig-g_1D}
\end{center}
\end{figure}

Starting from the 3D s-wave scattering length $a_\text{3D}$~[\onlinecite{Zuern13}], we calculate $g_\text{1D}$ as:
\begin{equation}
g_\text{1D}=\frac{2 \hbar^2 a_\text{3D}}{m a_\perp^2} \frac{1}{1-C a_\text{3D}/\sqrt{2} a_\perp} ,
\label{eq-gOneD}
\end{equation}
where $m$ is the mass of a $^6\text{Li}$ atom, ${a_\perp = \sqrt{\frac{\hbar}{m \omega_\perp}}}$ is the harmonic oscillator length of the radial confinement and ${C=-\zeta(\frac{1}{2})\approx1.46}$ with the Riemann zeta function $\zeta$~[\onlinecite{Olshanii98}]. Figure~\ref{fig-g_1D} shows the calculated 1D coupling constant $g_\text{1D}$ between two atoms in the two lowest hyperfine states of $^6\text{Li}$, which are trapped in our optical dipole trap with trap depth parameter $p=1$. All values of $g_\text{1D}$ are given in units of $a_{||} \hbar \omega_{||}$, where $a_{||} = \sqrt{\frac{\hbar}{m \omega_{||}}}$ is the harmonic oscillator length in longitudinal direction. In the tilted potential used for the tunneling measurements [Fig.~1(c) of the main text], we calculate $\omega_{||}$ by expanding the combined potential $V_\text{optical}(z)+V_\text{magnetic}(z)$ (Eq.~\ref{dipole_trap} and Eq.~\ref{magnetic_gradient}) around its minimum. 

\section{V. \quad	Measurement of occupation-number distributions}
\label{sec:occ_num}

\noindent To measure the mean occupation numbers on single-particle trap levels, we perform a series of measurements where we spill all population above a certain trap level $i$ from the trap (Section~II) and count the number of remaining atoms (Section~III). The mean number $N^{(i)}$ of remaining atoms corresponds to the sum of the populations on trap levels 0 to $i$. For the ground state, the mean occupancy is directly given by $N^{(0)}$. For excited states, we obtain the mean occupancies by subtracting $N^{(i-1)}$ from $N^{(i)}$.

\begin{figure}[t]
\begin{center}
\includegraphics[width = 0.5\columnwidth]{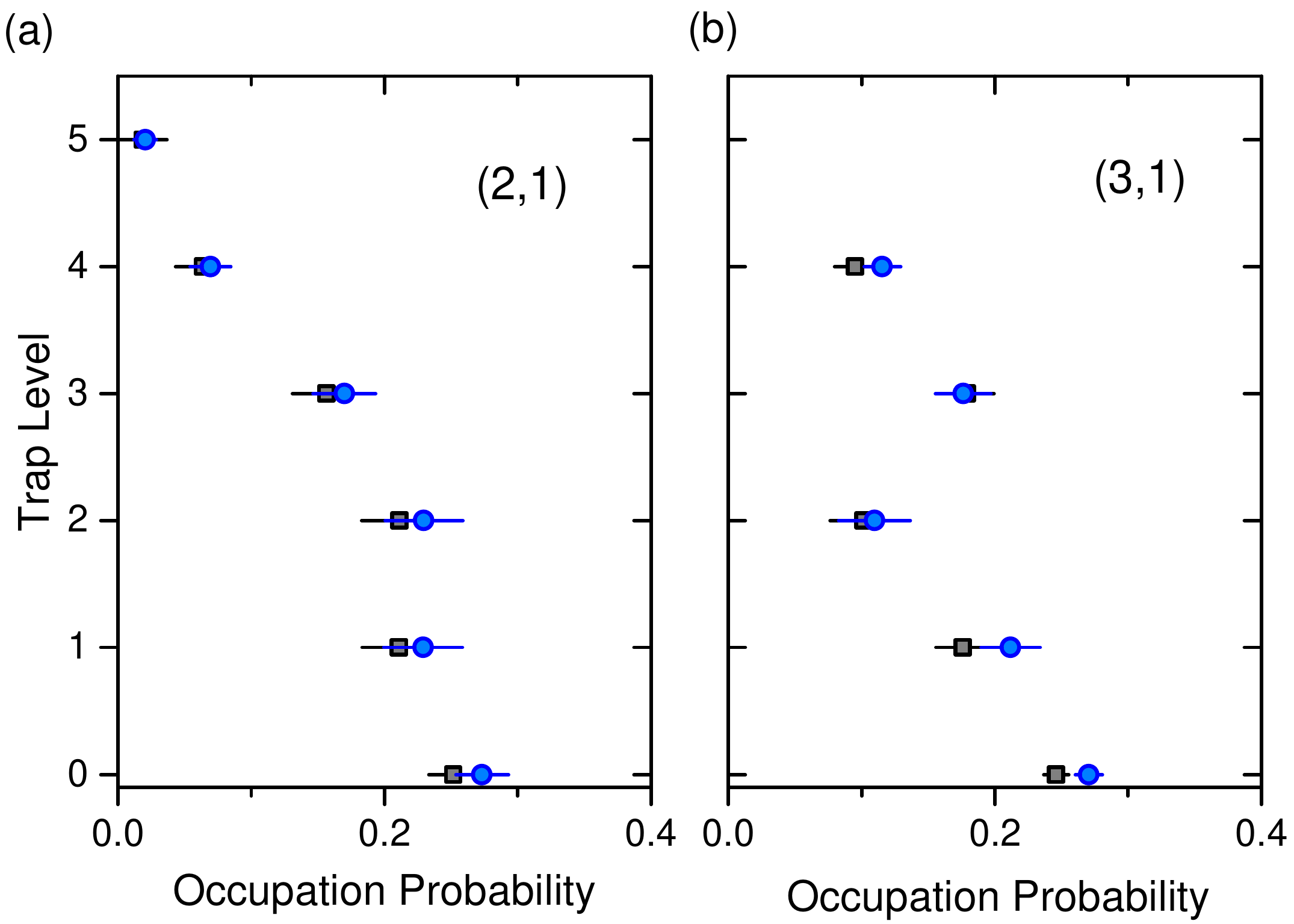}
\caption{Correction of occupation-number distributions. Corrected (blue circles) and uncorrected (black squares) occupation probabilities of the spin-down atom on single-particle trap levels for a $(2,1)$ system (a) and a $(3,1)$ system (b). For details on the measurement see Fig.~4 of the main text. Error bars denote the $1 \sigma$ statistical uncertainties.}
\label{fig-occupation_correction}
\end{center}
\end{figure}

When measuring the occupation-number distribution of a single spin-down atom in the trap (Fig.~4 of the main text), we correct for the finite fidelity of our experiment. During each experimental run, we measure the probability $P_{\text{zero}}$ to detect zero instead of one spin-down atom in a trap that still contains many trap levels. In an ideal experiment, we would expect this probability to be zero. Due to the finite fidelity of both the preparation of the initial state and the detection of the atom number, we measure values of $P_{\text{zero}} \approx 10~\%$ in all experimental runs. We divide each measured mean atom number $N^{(i)}$ by $(1 - P_{\text{zero}})$ to correct the data for this finite experimental fidelity. Figure~\ref{fig-occupation_correction} shows the effect of this correction for the data of Fig.~4 of the main text.

\section{VI. \quad	Confirmation that final state is ferromagnetic}
\label{sec-FM-final-state}

\begin{figure}[h]
\begin{center}
\includegraphics[width = 0.5\columnwidth]{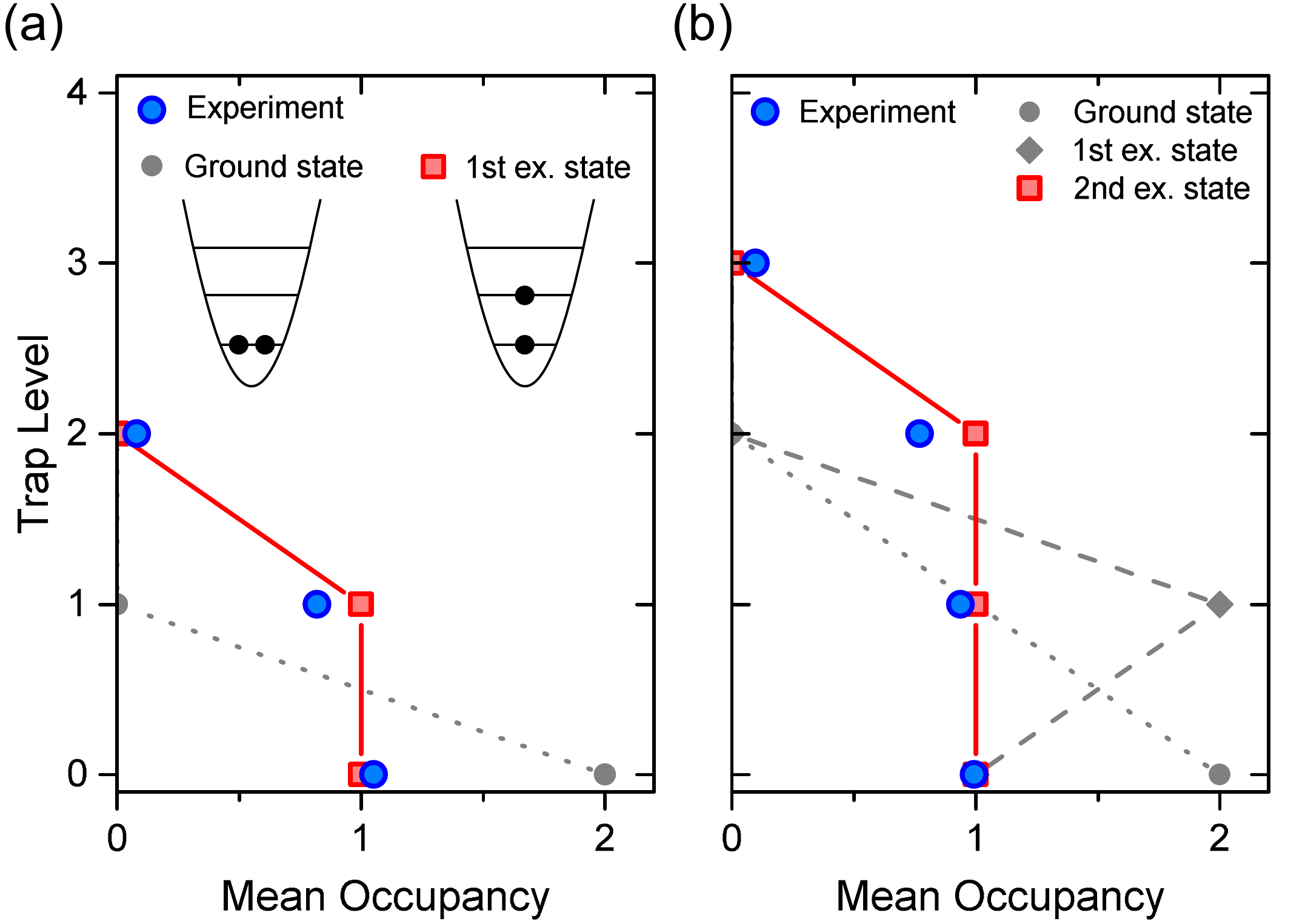}
\caption{Ferromagnetic final states. Combined occupation-number distribution of both spin-up and spin-down atoms on single-particle trap levels after tunneling of one atom in the super-Tonks regime. (a), The final two-atom state of an initial $(2,1)$ system after tunneling of one atom at $-1/g_\text{1D}=0.495\pm0.015$. (b), The final three-atom state of an initial $(3,1)$ system after tunneling of one atom at $-1/g_\text{1D}=0.553\pm0.015$. In both cases, the occupation-number distribution was measured after ramping the coupling constant back to $g_\text{1D}=0$. The red and gray symbols show the occupation-number distributions of the noninteracting states that are adiabatically connected to eigenstates of the ground-state multiplet in the spin-chain regime. In particular the red symbols correspond to the expectation for a ferromagnetic state. Error bars denoting the $1 \sigma$ statistical uncertainties are smaller than the symbols.}
\label{fig-FM_state}
\end{center}
\end{figure}

\noindent Our tunneling model predicts that above the CIR tunneling into the FM final states becomes predominant. To confirm this prediction, we show that both the initial $(2,1)$ and $(3,1)$ systems end up in a FM $(N-1)$-particle state after the tunneling of one particle deep in the super-Tonks regime. For the initial $(3,1)$ [$(2,1)$] system, the tunneling process is performed at an interactions strength of $-1/g_\text{1D}=0.553\pm0.015$ [$-1/g_\text{1D}=0.495\pm0.015$] and a trap depth parameter $p=0.846\pm0.025$ [$p=0.759\pm0.023$]. We ramp the final $(N-1)$-particle system back through the fermionization regime to zero interaction strength. Then, we measure the occupation number distributions on the few lowest single-particle trap levels (Section~V) as shown in Fig.~\ref{fig-FM_state}. Within our experimental fidelity, both distributions coincide with the expectation for a FM state. Note that in this measurement, the trap for the $(3,1)$ system was significantly deeper than in the measurements for Fig.~3(b) of the main text (See Section~VII). This could explain the deviation between the data of Fig.~3(b) of the main text and the expected limiting value of $P_\downarrow=0.75$ in the super-Tonks regime.

\begin{figure}[h]
\begin{center}
\includegraphics[width = 0.5\columnwidth]{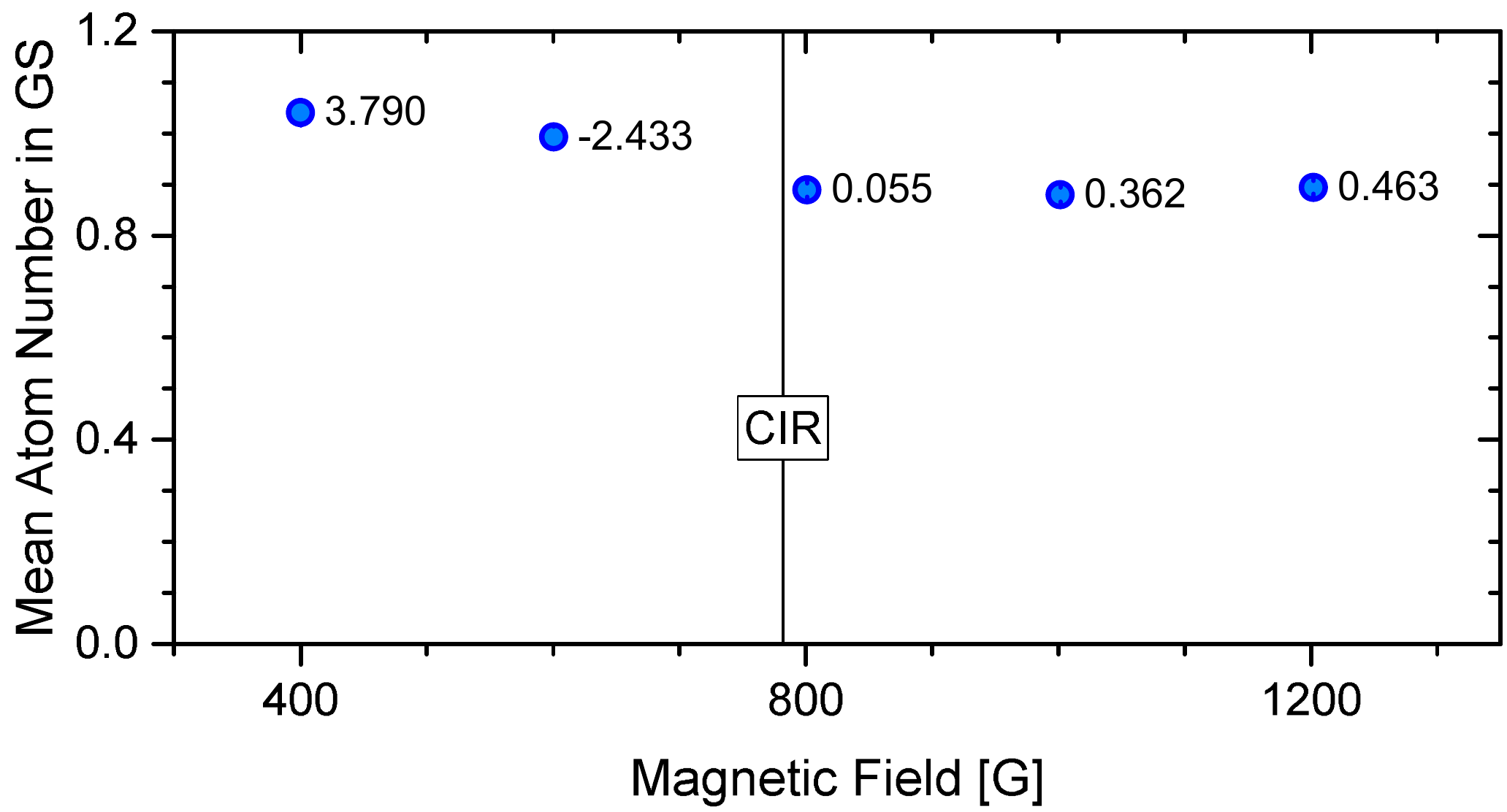}
\caption{Confirmation that the FM state is noninteracting. Ground-state population of the final state after tunneling of one atom from a $(2,1)$ system at $-1/g_\text{1D}=0.495\pm0.015$ as a function of magnetic field. The labels show the inverse 1D coupling constant $-1/g_\text{1D}$ at each point in units of $[a_{||}\hbar\omega_{||}]^{-1}$. The small increase of the ground-state population at $600~G$ and $400~G$ can be explained by the reduced magnetic moment of $^6\text{Li}$ at low magnetic fields, leading to a finite probability of detecting the population of the first excited state. Error bars denoting the $1 \sigma$ statistical uncertainties are smaller than the symbols.}
\label{fig-GS_occupation}
\end{center}
\end{figure}

The spatial wave function of a system of fermions in a ferromagnetic state is completely antisymmetric, which prohibits any two particles to be at the same place. Hence, a system of ultracold fermionic atoms with contact interactions in a FM state has to be noninteracting. We confirm this for the final two-particle state after letting one atom tunnel from the $(2,1)$ system above the CIR by measuring that independent of the interaction strength always one atom occupies the ground state of the trap (Fig.~\ref{fig-GS_occupation}).

\begin{figure}[ht]
\begin{center}
\includegraphics[width = 0.85\columnwidth]{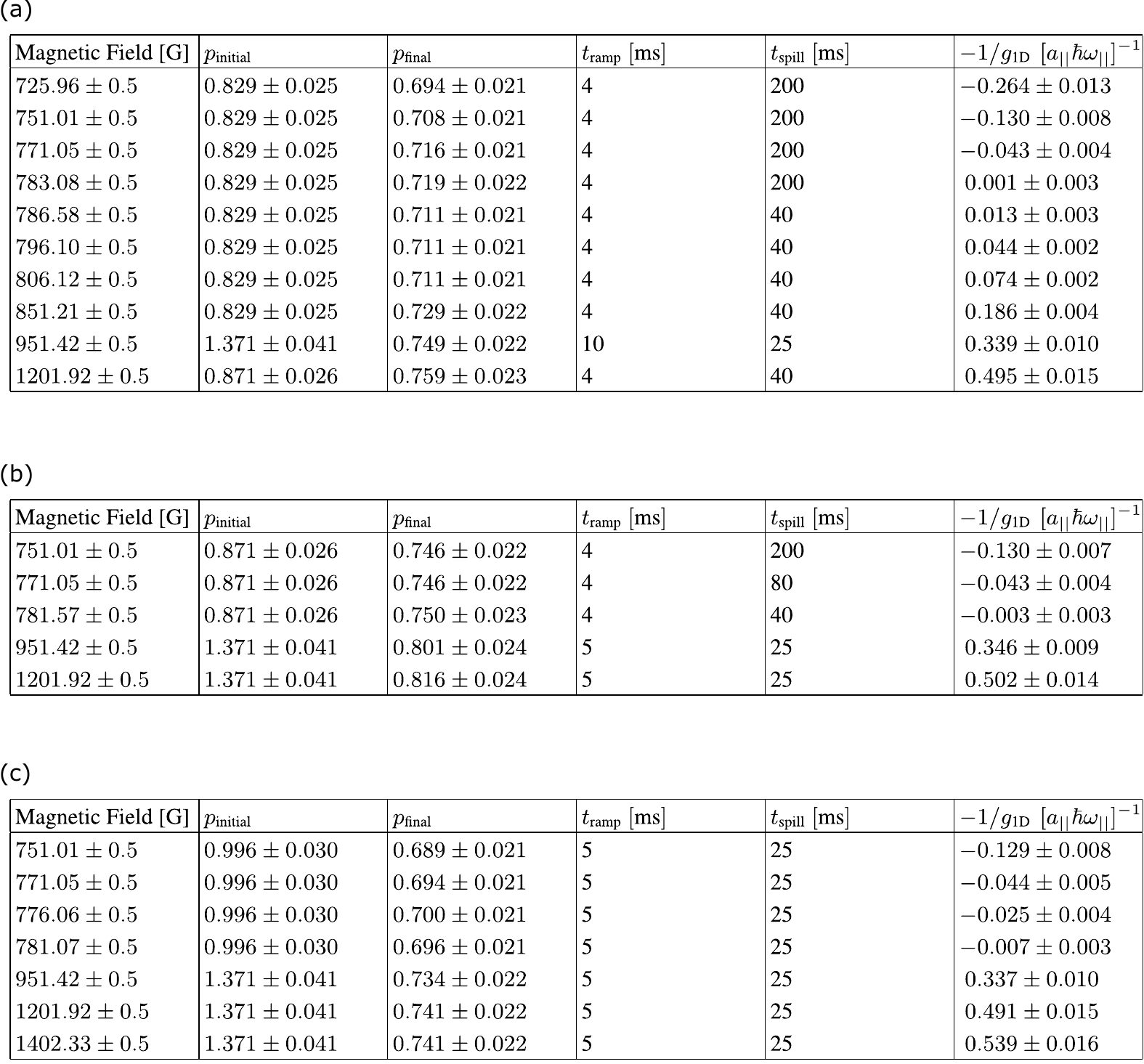}
\caption{Trap parameters during tunneling measurement. Magnetic fields, initial and final trap depth parameters, ramp times, spill times, and inverse 1D coupling constants for the measurement of spin-down tunneling (Fig.~3 of main text) for an initial $(2,1)$ system (a), a $(3,1)$ system (b), or a $(2,2)$ system (c). The final trap depth parameter $p_\text{final}$ during the tunneling process is optimized to let exactly one atom [in (a), and (b)] or two atoms [in (c)] tunnel out of the trap. Each row corresponds to one data point in Fig.~3 of the main text.}
\label{fig-spin_down_table}
\end{center}
\end{figure}

\section{VII. \quad	Measurement of probabilities of spin-down tunneling}

\label{sec-measurement-of-Pdown}

\noindent To measure the probability of spin-down tunneling (Fig.~3 of the main text), we let a specific number of atoms (one atom in the $(2,1)$ and the $(3,1)$ system, two atoms in the $(2,2)$ system) tunnel from the trap and measure the probability of spin-down tunneling. To do this, we superimpose the optical dipole trap (Eq.~\ref{dipole_trap}) with a magnetic field gradient (Eq.~\ref{magnetic_gradient}) to tilt the trap. Thereby we create a potential barrier between the in-trap states and the continuum of states outside the trap. To initiate the tunneling process, we ramp down the intensity of the trapping light within a time $t_\text{ramp}$ and thereby lower the trap depth parameter $p$ (Eq.~\ref{dipole_trap}) from its initial value $p_\text{initial}$ to a final value $p_\text{final}$. After a time $t_\text{spill}$, during which atoms can tunnel out of the trap, we ramp the power back to its original value to stop the tunneling process. In Table~\ref{fig-spin_down_table}a, b, and c the parameters for the tunneling measurements on the $(2,1)$, the $(3,1)$, and the $(2,2)$ system are listed. For the $(2,1)$ and the $(3,1)$ systems, the parameters lead to the tunneling of exactly one atom, while for the $(2,2)$ system two atoms tunnel out of the trap.

\begin{figure}
\begin{center}
\includegraphics[width = 0.9\columnwidth]{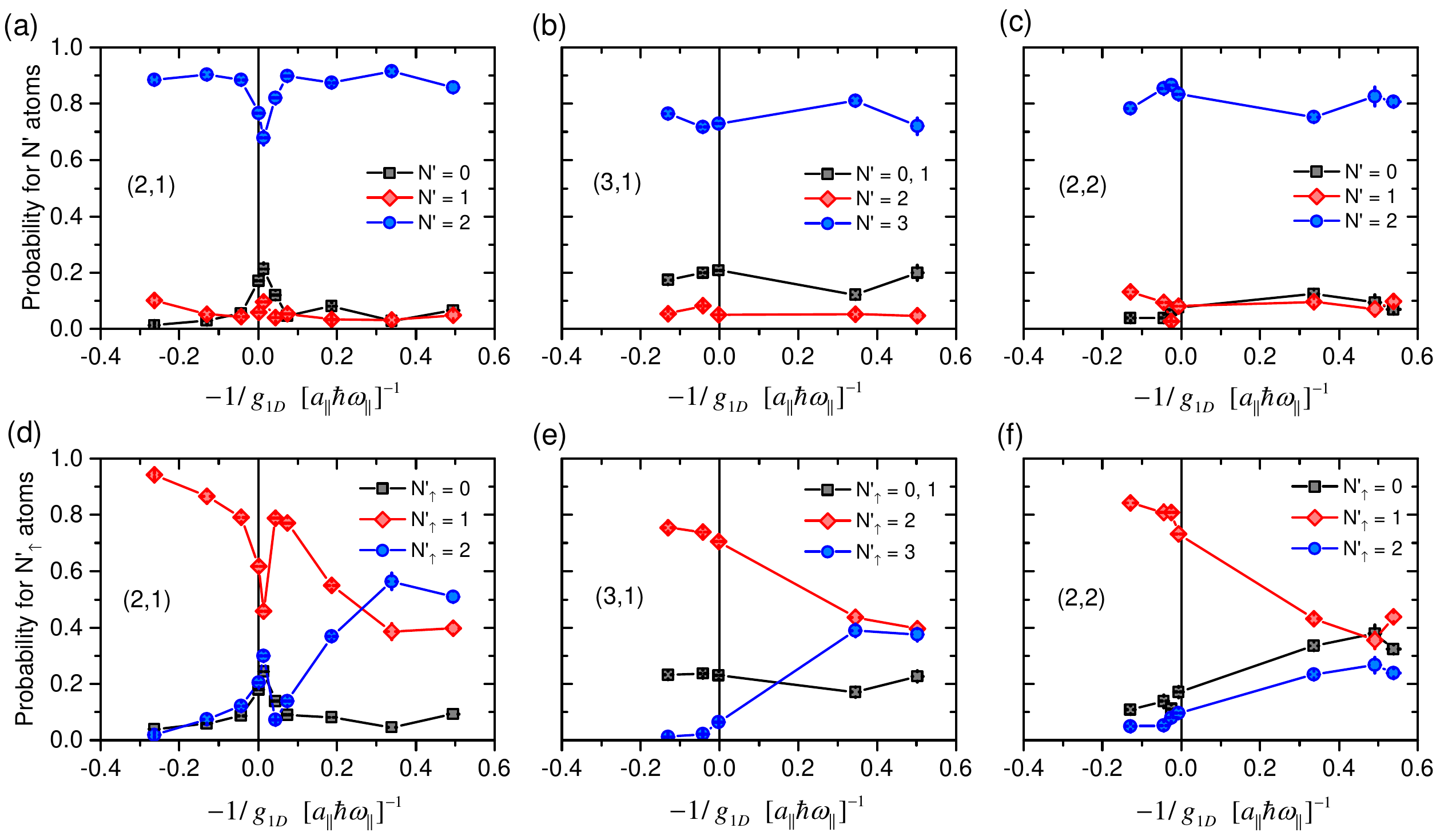}
\caption{Raw data of spin-down tunneling. Probabilities to measure $N'$ atoms [(a), (b), (c)] or $N'_\uparrow$ spin-up atoms [(d), (e), (f)] after the tunneling of one atom [(a), (b), (d), (e)], or of two atoms [(c), (f)] at different interaction strenghts. (a) and (d) are for an initial $(2,1)$ system. (b) and (e) are for an initial $(3,1)$ system. (c) and (f) are for an initial $(2,2)$ system. Error bars denote the $1 \sigma$ statistical uncertainties.}
\label{fig-spin_down_raw}
\end{center}
\end{figure}

Throughout the experimental runs, we perform control measurements in which we detect the number of atoms ($N'$) directly after the tunneling process (see Fig.~\ref{fig-spin_down_raw}(a), (b), and (c) for the control measurements on the $(2,1)$, the $(3,1)$, and the $(2,2)$ system). The blue circles show the probability to measure the expected $N'$ (2 atoms for the $(2,1)$ system, 3 atoms for the $(3,1)$ system, and 2 atoms for the $(2,2)$ system). This probability is always smaller than one due to the finite experimental fidelity of our measurements and due to three and four-body losses. Since our experimental fidelity is on the order of $90\%$, we can assume that shots with two or more atoms missing [black squares in Fig.~\ref{fig-spin_down_raw}(a), (b), (c)] are primarily a result of three and four-body losses. For the $(2,1)$ system, three-body losses mostly appear for tunneling close to the resonance [black squares in Fig.~\ref{fig-spin_down_raw}(a)], while for the $(3,1)$ and the $(2,2)$ systems trap losses can be observed for tunneling at all interaction strengths [black squares in Fig.~\ref{fig-spin_down_raw}(b), (c)].

After the initial tunneling process, we selectively remove the spin-down component from the trap and measure the number of spin-up atoms ($N_\uparrow'$) in the trap [Fig.~\ref{fig-spin_down_raw}(d), (e), and (f) for the $(2,1)$, the $(3,1)$ and the $(2,2)$ systems]. In these figures, the blue circles correspond to the probability of spin-down tunneling.

\begin{figure}
\begin{center}
\includegraphics[width = 0.9\columnwidth]{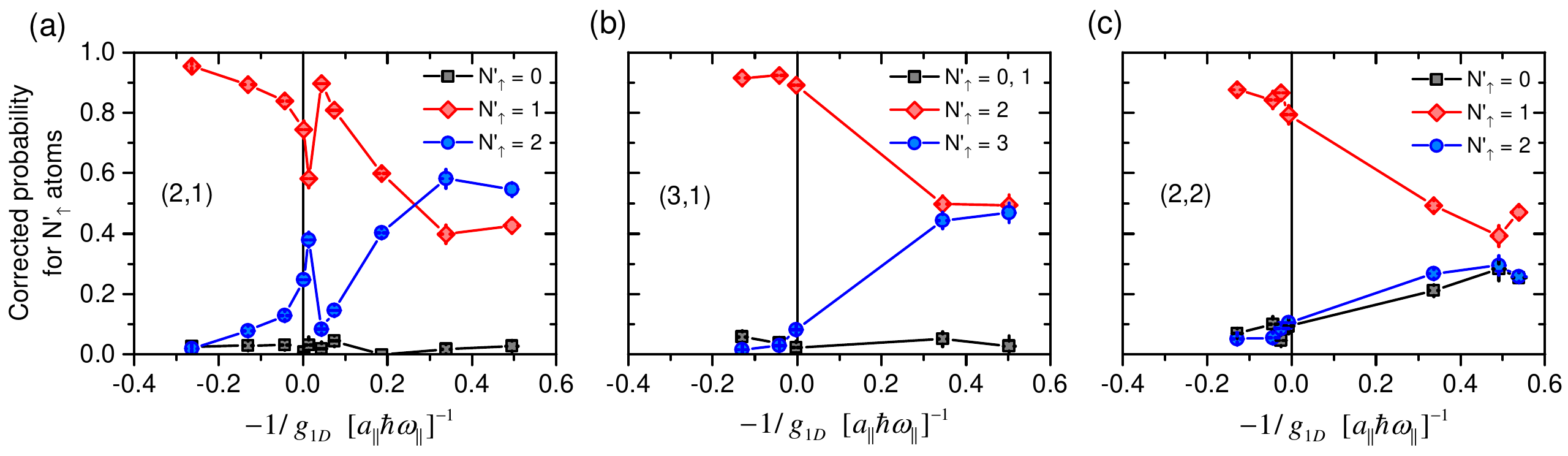}
\caption{Corrected data of spin-down tunneling. Corrected probabilities of the number of spin-up atoms $N'_\uparrow$ after the tunneling of one atom [(a), (b)] or of two atoms (c) at different interaction strengths. (a) is for an initial $(2,1)$ system. (b) is for an initial $(3,1)$ system. (c) is for an initial $(2,2)$ system. The blue circles in (a) and (b) correspond to the probability of tunneling of the spin-down atom. The blue circles in (c) correspond to the probability of tunneling of both spin-down atoms. Error bars denote the $1 \sigma$ statistical uncertainties.}
\label{fig-spin_down_correction}
\end{center}
\end{figure}

We correct the probability of spin-down tunneling for three and four-body losses, by subtracting the probabilities of trap losses [black squares in Fig.~\ref{fig-spin_down_raw}(a), (b), (c)] from the respective probabilities to detect 0 spin-up atoms (0, or 1 spin-up atoms in the $(3,1)$ system) [black squares in Fig.~\ref{fig-spin_down_raw}(d), (e), (f)]. Since the sum of all probabilities has to be normalized, this correction increases the values of both the red and blue symbols in Fig.~\ref{fig-spin_down_raw}(d), (e), and (f), while keeping their respective ratio constant. The corrected probabilities are shown in Fig.~\ref{fig-spin_down_correction}. The blue circles in this figure are the corrected probabilities of spin-down tunneling that are also shown in Fig.~3 of the main text.

\section{VIII. \quad	Theoretical model}

\subsection{Spin-chain model}

In the fermionization regime of diverging interaction strength, a 1D multicomponent system of $N$ fermionic atoms acquires the density distribution of $N$ identical noninteracting fermions. In this limit, the ordering of the atoms is fixed and a one-to-one correspondence between the full wave function of space and spin and a pure spin wave function can be established~[\onlinecite{Deuretzbacher08, Deuretzbacher14}]. This implies that all observables can be mapped from the full continuous Hilbert space into a discrete spin space and vice versa. This mapping therefore allows to treat a 1D system in the fermionization regime with a spin-chain model and therefore constitutes a substantial simplification of its theoretical description.

For a spin state $\left|\chi\right\rangle$ consisting of $N$ spin-1/2 particles, we will use the notation
\begin{equation}
\left|\chi\right\rangle = \left|m_1\right\rangle \otimes ... \otimes \left|m_N\right\rangle \equiv \left|m_1,...,m_N\right\rangle,
\end{equation}
where the spin state $\left|m_i\right\rangle$ of the $i$th particle in the spin chain can either be up $\big(\left|\uparrow\right\rangle\big)$ or down  $\big(\left|\downarrow\right\rangle\big)$.

The spin-chain Hamiltonian of $N$ spins is given by~[\onlinecite{Deuretzbacher14}]
\begin{equation}
H_s^{(N)} = \left( E_F^{(N)} - \sum_{i=1}^{N-1} J_i \right) \openone + \sum_{i=1}^{N-1} J_i P_{i,i+1} .
\label{eq:spin-chain_hamiltonian}
\end{equation}
Here, $E_F^{(N)}$ is the energy of $N$ spinless noninteracting fermions in 1D, $i$ refers to the $i$th particle, $P_{i,i+1}$ permutes neighboring spins and $J_i$ is the exchange coupling between neighboring spins. $J_i$  is given by the formula
\begin{equation}
J_i = \frac{N! \hbar^4}{m^2 g_\text{1D}} \int dz_1 \dotsi dz_N \delta (z_i-z_{i+1}) \theta (z_1, \dotsc, z_N) \left| \frac{\partial \psi_F}{\partial z_i} \right|^2 .
\end{equation}
Here, $\psi_F$ is the ground state of $N$ spinless noninteracting fermions in 1D and the function $\theta (z_1, \dotsc , z_N)$ is 1 if ${z_1 \leq \dotsm \leq z_N}$ and zero otherwise. It is important to note that $J_i$ is proportional to the inverse interaction strength ($1/g_\text{1D}$) and that it is approximately proportional to the local density cubed and therefore depends on the trap geometry.

To determine the exchange couplings $J_i$ for our tunneling measurements, we calculate the quasi-bound states $\psi_F$ in the tilted trap (Fig.~1c of main text) as follows: First, we approximate the potential by the Taylor series around its local minimum. We adjust the maximum order of the Taylor series such that the potential is well approximated up to the barrier maximum and that the Taylor series is larger than the barrier maximum outside the quasi-bound region. Then, we calculate the spectrum and the eigenfunctions of the approximate potential using the eigenfunctions of the harmonic oscillator at the local minimum.

\subsection{Tunneling model}

In our model, we consider tunneling of one atom from the trap. Hence, the model connects an initial spin state $|i\rangle$, which is an eigenstate of $H_s^{(N)}$, with a final spin state ${|f,t\rangle \equiv |f\rangle \otimes |t\rangle}$. In the final state, $|f\rangle$ is an eigenstate of $H_s^{(N-1)}$ and describes the $N-1$ particles that remain in the trap and $|t\rangle$ indicates the spin orientation of the particle that left the trap.

The energy $E_{i,f}$ of the tunneling particle is equal to the difference between the energies of the initial and final in-trap states. Close to the CIR, the energy shifts of all states are linear in $-1/g_\text{1D}$ (Fig.~2 of the main text) and the eigenenergies of $|i\rangle$ and $|f\rangle$ are well approximated by the solutions of the spin-chain model. In this limit, their difference is
\begin{equation}
E_{i,f} = \langle i|H_s^{(N)}|i \rangle - \langle f|H_s^{(N-1)}|f\rangle .
\end{equation}
Further away from the CIR, we interpolate between the linear energy shifts of the spin-chain model and the corresponding saturation energies at $-1/g_\text{1D} = \pm \infty$ using the rescaled energy shifts of the $(1,1)$ system in the ground state~[\onlinecite{Gharashi15}].

The weight of each tunneling channel $T_{i,f}$ for tunneling from $|i\rangle$ to $|f,t\rangle$ is approximated by
\begin{equation}
T_{i,f} \propto |\langle i|f,t\rangle|^2 E_{i,f} e^{-2\gamma (E_{i,f})} .
\label{eq:tunnelrate}
\end{equation}
That means that it is proportional to the squared spin overlap \big($|\langle i|f,t\rangle|^2$\big), the energy of the tunneling particle \big($E_{i,f}$\big) and the probability \big($e^{-2\gamma (E_{i,f})}$\big) that a particle with energy $E_{i,f}$ tunnels through the barrier. The energy-dependent tunneling parameter $\gamma (E_{i,f})$ is given by
\begin{equation}
\gamma(E_{i,f}) = \frac{1}{\hbar} \int_{z_1}^{z_2} dz \sqrt{2m \bigl[ V(z)-E_{i,f} \bigr]} .
\end{equation}
Here, $z_1$ and $z_2$ are the two largest solutions of $V(z)-E_{i,f}=0$.

The probability to tunnel from state $|i\rangle$ to state $|f\rangle$ is given by
\begin{equation}
P_{i,f} = \frac{T_{i,f}}{\left( \sum_{f'} T_{i,f'} \right)} ,
\label{eq:tunnelprob}
\end{equation}
where the sum is over all possible final states $|f'\rangle$.

Exactly on resonance, $-1/g_\text{1D}=0$, the energy differences $E_{i,f}$ are equal for all final states (Fig.~2 of the main text). Therefore, the energy-dependent term $E_{i,f} e^{-2\gamma (E_{i,f})}$ in Eq.~\ref{eq:tunnelrate} is equal for all tunneling channels and can be canceled from Eq.~\ref{eq:tunnelprob}. In this case the probability to tunnel from state $|i\rangle$ to state $|f,t\rangle$ simplifies to $P_{i,f} = |\langle i|f,t\rangle|^2$ since $\sum_{f'} |\langle i|f',t'\rangle|^2 = 1$. The limiting behavior of $P_{i,f}$ far left and right from the CIR is determined by the predominant tunneling channels of these regimes. For small repulsive interactions (left blue arrow in Fig.~2 of the main text), tunneling into the antiferromagnetic $(N-1)$-particle in-trap state is predominant since its energy is much smaller than the energies of all other $(N-1)$-particle states. For small attractive interactions (right blue arrow in Fig.~2 of the main text), tunneling into the FM $(N-1)$-particle in-trap state is predominant since the ordering of the energy levels is inverted. This is also confirmed by the measurements of Section~VI.

\section{IX. \quad	Calculation of spin-down tunneling probabilities}

Using Eq.~\ref{eq:tunnelprob}, we calculate the probabilities for spin-down tunneling $P_\downarrow(-1/g_\text{1D})$ as a function of interaction strength. Here, $P_\downarrow(-1/g_\text{1D})$ is the probability of tunneling from an initial state $|i\rangle$ into the final state $|f\rangle \otimes |t\rangle$, where the in-trap state $|f\rangle$ only contains spin-up atoms and $|t\rangle$ only contains spin-down atoms. The initial and final in-trap states ($|i\rangle$ and $|f\rangle$) of our tunneling model are always eigenstates of the spin-chain Hamiltonian (Eq.~\ref{eq:spin-chain_hamiltonian}). In particular, the initial state will always be the antiferromagnetic ground state $|AFM\rangle$ of the repulsive spin-chain model with $N$ atoms.

We will solve the $(2,1)$ and $(2,2)$ systems analytically to make the physics more transparent. To achieve this, we will first determine the eigenstates of the $(2,1)$ system in a symmetric trap and then show how they mix in a tilted trap (See Fig.~1(c) of the main text and Section~VII). In the $(2,2)$ case, we will first determine the two eigenstates of $\vec S^2$ with $S=0$. Again, we will calculate the spin-chain Hamiltonian within the subspace of these two states and show how they couple in a tilted trap. We will show that both systems are closely related, since the $(2,2)$ system ends up in the AFM states of the three-particle system with almost 100\% probability after the first tunneling process.

\subsection{(2,1) system}

Within the spin basis ${|1\rangle = | \uparrow, \uparrow, \downarrow \rangle}$, ${|2\rangle = | \uparrow, \downarrow, \uparrow \rangle}$, and ${|3\rangle = | \downarrow, \uparrow, \uparrow \rangle}$ the spin-chain Hamiltonian reads
\begin{equation}
H_s^{(3)} = E_F^{(3)} \openone +
\begin{pmatrix}
  -J_2 & J_2 & 0 \\
  J_2 & -J_1-J_2 & J_1 \\
  0 & J_1 & -J_1
\end{pmatrix} .
\end{equation}
It has in a symmetric trap ($J_1=J_2$) the eigenstates~[\onlinecite{Deuretzbacher14}]
\begin{equation}
| AFM_3 \rangle = \frac{1}{\sqrt{6}} \Bigl( | \uparrow , \uparrow , \downarrow \rangle - 2 | \uparrow , \downarrow , \uparrow \rangle + | \downarrow , \uparrow , \uparrow \rangle \Bigr) ,
\end{equation}
\begin{equation}
| IM_3 \rangle = \frac{1}{\sqrt{2}} \Bigl( | \uparrow , \uparrow , \downarrow \rangle - | \downarrow , \uparrow , \uparrow \rangle \Bigr) ,
\end{equation}
\begin{equation} \label{eq-FM}
| FM_3 \rangle = \frac{1}{\sqrt{3}} \Bigl( | \uparrow , \uparrow , \downarrow \rangle + | \uparrow , \downarrow , \uparrow \rangle + | \downarrow , \uparrow , \uparrow \rangle \Bigr)
\end{equation}
and the eigenenergies $E_{AFM}^{(3)} = E_F^{(3)} - 3 J_1$, $E_{IM}^{(3)} = E_F^{(3)} - J_1$, $E_{FM}^{(3)} = E_F^{(3)}$. In this eigenbasis of the symmetric trap, the spin-chain Hamiltonian reads
\begin{equation}
H_s^{(3)} = E_F^{(3)} \openone +
\begin{bmatrix}
  -\frac{3}{2}(J_1+J_2) & \frac{\sqrt{3}}{2}(J_1-J_2) & 0 \\
  \frac{\sqrt{3}}{2}(J_1-J_2) & -\frac{1}{2}(J_1+J_2) & 0 \\
  0 & 0 & 0
\end{bmatrix} .
\end{equation}
Therefore, an imbalance of $J_1$ and $J_2$ coherently mixes the AFM with the IM state, while the FM state is decoupled. This can be understood, since both the AFM and the IM state have an absolute value of the total spin of $S=1/2$, while the FM state has $S=3/2$. The eigenstates of $H_s^{(3)}$ in a nonharmonic trap can hence be written as
\begin{equation} \label{eq-AFM}
| AFM'_3 \rangle = \cos \frac{\alpha}{2} | AFM_3 \rangle - \sin \frac{\alpha}{2} | IM_3 \rangle ,
\end{equation}
\begin{equation} \label{eq-IM}
| IM'_3 \rangle = \cos \frac{\alpha}{2} | IM_3 \rangle + \sin \frac{\alpha}{2} | AFM_3 \rangle ,
\end{equation}
and its eigenenergies are $E_{AFM'}^{(3)} = E_F^{(3)} - (J_1+J_2) \left( 1 + \frac{1}{2 \cos \alpha} \right)$, $E_{IM'}^{(3)} = E_F^{(3)} - (J_1+J_2) \left( 1 - \frac{1}{2 \cos \alpha} \right)$ with the mixing angle $\alpha = \arctan \left( \sqrt{3} \frac{J_1-J_2}{J_1+J_2} \right)$.

The $(2,1)$ system is initially prepared in state $| AFM'_3 \rangle$. Exactly on resonance, $-1/g_\text{1D}=0$, the ordering of the particles is fixed by their mutual repulsion such that only the rightmost particle can leave the trap (Fig.~1(c) of the main text). The probability for spin-down tunneling, $P_\downarrow$, is hence just the probability of the rightmost spin of the $| AFM'_3 \rangle$ state to point downwards,
\begin{equation}
P_\downarrow(-1/g_\text{1D}=0) = |\langle \uparrow , \uparrow , \downarrow | AFM'_3 \rangle|^2 = \left( \cos \frac{\alpha}{2} \frac{1}{\sqrt{6}} - \sin \frac{\alpha}{2} \frac{1}{\sqrt{2}} \right)^2 .
\end{equation}
In the symmetric trap $(\alpha = 0)$, the probability for spin-down tunneling is $P_\downarrow(-1/g_\text{1D}=0) = 1/6 = 16.7\%$. In the tilted trap, $\alpha \propto J_1-J_2$ is positive, since the density between the left particle and the particle in the middle is larger than the density between the particle in the middle and the right particle (Fig.~1(c) of the main text). $P_\downarrow(-1/g_\text{1D}=0)$ is hence lower than $16.7\%$ in the tilted trap. We obtain for our experimental trap parameters during the tunneling process a mixing angle of $\alpha \approx 20^\circ$ resulting in $P_\downarrow(-1/g_\text{1D}=0) \approx 8\%$ for the $| AFM'_3 \rangle$ state (Fig.~3(a) of the main text).

Away from resonance, $-1/g_\text{1D} \neq 0$, one has to calculate the weights of all the possible tunneling channels. The eigenstates of $H_s^{(2)}$ and $S_z^{(3)}$ are $| FM_{2,1} \rangle = | \uparrow , \uparrow , \downarrow \rangle$, $| FM_{2,2} \rangle = \frac{1}{\sqrt{2}} ( | \uparrow , \downarrow \rangle + | \downarrow , \uparrow \rangle ) | \uparrow \rangle$, and $| AFM_2 \rangle = \frac{1}{\sqrt{2}} ( | \uparrow , \downarrow \rangle - | \downarrow , \uparrow \rangle ) | \uparrow \rangle$. These eigenstates have the eigenenergies $E_{FM}^{(2)} = E_F^{(2)}$ and $E_{AFM}^{(2)} = E_F^{(2)} - 2 J_1$ (now calculated with $J_1$ of the two-particle in-trap system). The probability for spin-down tunneling, $P_\downarrow$, is the probability to tunnel from the initial state $| AFM'_3 \rangle$ into the final state $| FM_{2,1} \rangle = | \uparrow , \uparrow , \downarrow \rangle$. Far left from the CIR, tunneling into $| AFM_2 \rangle$ is predominant and one obtains the limiting value
\begin{equation}
P_\downarrow(-1/g_\text{1D} \rightarrow -\infty) = 0 .
\end{equation}
Far right from the CIR, tunneling into $| FM_{2,1} \rangle$ and $| FM_{2,2} \rangle$ is predominant and one obtains the limiting value
\begin{equation}
P_\downarrow(-1/g_\text{1D} \rightarrow +\infty) = \frac{| \langle FM_{2,1} | AFM_3 \rangle |^2}{| \langle FM_{2,1} | AFM_3 \rangle |^2 + | \langle FM_{2,2} | AFM_3 \rangle |^2} = \frac{2}{3}
\end{equation}
(see Fig.~3(a) of the main text). Interestingly, this limiting value is independent of the mixing angle $\alpha$ and hence the tilting of the trap. In between, $P_\downarrow = P_{| AFM'_3 \rangle, | \uparrow , \uparrow \rangle}$ is calculated using the parameters $p_\text{final}$ and $-1/g_\text{1D}$ of Table~\ref{fig-spin_down_table}(a). This leads to the red solid curve in Fig.~3(a) of the main text. Similar calculations for the IM and FM initial states lead to the dotted and dashed curves of Fig.~3(a) of the main text, respectively.

We assumed for the above calculations that the trap is tilted so slowly that the spin configuration can follow the change of $J_1$ and $J_2$ induced by the change of the total density. This assumption is wrong close to the CIR, where the spin configuration of the initially deeper trap may be frozen. We would hence expect a larger $P_\downarrow$ close to the CIR. A full dynamical calculation showed, however, that this effect is not sufficiently strong to explain the resonance feature of $P_\downarrow$ close to the CIR.

\subsection{(3,1) system}

Within the spin basis ${|1\rangle = | \uparrow, \uparrow, \uparrow, \downarrow \rangle}$, ${|2\rangle = | \uparrow, \uparrow, \downarrow, \uparrow \rangle}$, ${|3\rangle = | \uparrow, \downarrow, \uparrow, \uparrow \rangle}$, and ${|4\rangle = | \downarrow, \uparrow, \uparrow, \uparrow \rangle}$ the spin-chain Hamiltonian reads
\begin{equation}
H_s^{(4)} = E_F^{(4)} \openone +
\begin{pmatrix}
  -J_3 & J_3 & 0 & 0 \\
  J_3 & -J_2-J_3 & J_2 & 0 \\
  0 & J_2 & -J_1-J_2 & J_1 \\
  0 & 0 & J_1 & -J_1
\end{pmatrix} .
\end{equation}
The $(3,1)$ system is initially prepared in $| AFM_4 \rangle$, which is the ground state of this Hamiltonian for $g>0$. We determine this state numerically. We have to consider tunneling into the final states $| FM_{3,1} \rangle = | \uparrow, \uparrow, \uparrow, \downarrow \rangle$, $| FM_{3,2} \rangle = | FM_3 \rangle | \uparrow \rangle$, $| IM_3'' \rangle = | IM_3' \rangle | \uparrow \rangle$, and $| AFM_3'' \rangle = | AFM_3' \rangle | \uparrow \rangle$, where $| FM_3 \rangle$ is given by Eq.~(\ref{eq-FM}), $| IM_3' \rangle$ is given by Eq.~(\ref{eq-IM}), and $| AFM_3' \rangle$ is given by Eq.~(\ref{eq-AFM}). Exactly on resonance, the probability for spin-down tunneling is just given by
\begin{equation}
P_\downarrow(-1/g_\text{1D}=0) = |\langle \uparrow, \uparrow , \uparrow , \downarrow | AFM_4 \rangle|^2 .
\end{equation}
We obtain for the harmonic trap a value of $P_\downarrow(-1/g_\text{1D}=0) = 5.1\%$~[\onlinecite{Deuretzbacher14}]. This value is again lowered by the tilting and we obtain for the experimental parameters at the CIR (see Table~\ref{fig-spin_down_table}(b)) a value of $P_\downarrow(-1/g_\text{1D}=0) \approx 1\%$. Far left from the CIR, tunneling into $| AFM_3'' \rangle$ is predominant and one obtains again the limiting value $P_\downarrow(-1/g_\text{1D} \rightarrow -\infty) = 0$. Far right from the CIR, tunneling into the two FM final states, $| FM_{3,1} \rangle$ and $| FM_{3,2} \rangle$, is predominant and one obtains the limiting value
\begin{equation}
P_\downarrow(-1/g_\text{1D} \rightarrow +\infty) = \frac{| \langle FM_{3,1} | AFM_4 \rangle |^2}{| \langle FM_{3,1} | AFM_4 \rangle |^2 + | \langle FM_{3,2} | AFM_4 \rangle |^2} = \frac{3}{4} ,
\end{equation}
which is again independent of the tilting. The experimentally measured value for $P_\downarrow$ is, however, only $\approx 50\%$ in the super-Tonks regime (see Fig.~3(b) of the main text). The deviation from the theoretical prediction ($\approx 70\%$) is presumably caused by a stronger contribution of the IM tunneling channel in the experiment.
% Indeed, for a shallower trap ($p_\text{final}=0.735$), where the fourth trap level is only slightly below the barrier maximum, we calculate $P_\downarrow \approx 50\%$.
Indeed, the IM tunneling channel contributes stronger if the lowering of the potential barrier is assumed to be nonadiabatic, as outlined in the following paragraphs.

In the fermionization regime, the energy of the energetically highest particle is given by the $N$th trap level. This energy grows linearly with $-1/g_\text{1D}$ in the fermionization regime and it saturates in the weakly-interacting regimes, $-1/g_\text{1D} = \pm \infty$, at values, which are again given by the (quasi-bound) trap levels~[\onlinecite{Gharashi13}]. The saturated energy shifts can hence never exceed the barrier maximum. This situation applies to an adiabatic lowering of the potential barrier before the tunneling process. This condition is fulfilled, if the tilted trap is deep enough or if the barrier is lowered slowly when the corresponding trap level approaches the barrier maximum.

If the barrier height is nonadiabatically lowered, the saturated energy shifts may exceed the barrier maximum. In this case the saturated shifts are multiples of the level spacing at the Fermi edge and they are not limited by the barrier maximum. The energy of the tunneling particle may now exceed the barrier maximum for the energetically most favorable tunneling channels. In that case the particle tunnels out with a probability of 1. Moreover, the tunneling energies of the intermediate channels may also be slightly below or above the barrier maximum. This is the case for the IM tunneling channel of the $(3,1)$ system in the super-Tonks regime. As a result, the weight of the IM tunneling channel is much stronger than in the adiabatic approximation, which leads to a substantial reduction of $P_\downarrow$ in the super-Tonks regime. We obtain $P_\downarrow(-1/g_\text{1D} = 0.35) = 38\%$ and $P_\downarrow(-1/g_\text{1D} = 0.5) = 61\%$ for the $(3,1)$ system in the super-Tonks regime, which is in much better agreement with the experimental data. We finally note that, in contrast to the $(3,1)$ system, both models lead to nearly the same results for the $(2,1)$ and $(2,2)$ systems.

\subsection{(2,2) system}

Within the spin basis $|1\rangle = | \uparrow, \uparrow, \downarrow, \downarrow \rangle$, $|2\rangle = | \uparrow, \downarrow, \uparrow, \downarrow \rangle$, $|3\rangle = | \downarrow, \uparrow, \uparrow, \downarrow \rangle$, $|4\rangle = | \downarrow, \downarrow, \uparrow, \uparrow \rangle$, $|5\rangle = | \downarrow, \uparrow, \downarrow, \uparrow \rangle$, and $|6\rangle = | \uparrow, \downarrow, \downarrow, \uparrow \rangle$ the spin-chain Hamiltonian reads
\begin{equation}
H_s^{(4)} = E_F^{(4)} \openone +
\begin{pmatrix}
  -J_2 & J_2 & 0 & 0 & 0 & 0 \\
  J_2 & -J_1-J_2-J_3 & J_1 & 0 & 0 & J_3 \\
  0 & J_1 & -J_1-J_3 & 0 & J_3 & 0 \\
  0 & 0 & 0 & -J_2 & J_2 & 0 \\
  0 & 0 & J_3 & J_2 & -J_1-J_2-J_3 & J_1 \\
  0 & J_3 & 0 & 0 & J_1 & -J_1-J_3
\end{pmatrix} .
\end{equation}
The $(2,2)$ system is initially prepared in $| AFM_4' \rangle$, which is the ground state of this Hamiltonian for $g>0$. It can be determined analytically since only the ground state and another excited have an absolute value of the total spin of $S=0$.
% The $\vec S^2$ operator is in the above basis given by
% \begin{equation}
% \vec S^2 =
% \begin{pmatrix}
%   2 & 1 & 1 & 0 & 1 & 1 \\
%   1 & 2 & 1 & 1 & 0 & 1 \\
%   1 & 1 & 2 & 1 & 1 & 0 \\
%   0 & 1 & 1 & 2 & 1 & 1 \\
%   1 & 0 & 1 & 1 & 2 & 1 \\
%   1 & 1 & 0 & 1 & 1 & 2
% \end{pmatrix} .
% \end{equation}
The two eigenstates with $S=0$ are
\begin{equation}
| S=0, 1 \rangle = \frac{1}{2} \Bigl( |1\rangle - |3\rangle + |4\rangle - |6\rangle \Bigr)
\end{equation}
and
\begin{equation}
| S=0, 2 \rangle = \frac{1}{\sqrt{12}} \Bigl( |1\rangle - 2 |2\rangle + |3\rangle + |4\rangle - 2 |5\rangle + |6\rangle \Bigr) .
\end{equation}
In the subspace of these two states, the spin-chain Hamiltonian attains the form
\begin{equation}
\left. H_s^{(4)} \right|_{S=0} = E_F^{(4)} \openone +
\begin{bmatrix}
  -\frac{1}{2}(J_1+J_2+J_3) & \frac{\sqrt{3}}{2}(J_1-J_2+J_3) \\
  \frac{\sqrt{3}}{2}(J_1-J_2+J_3) & -\frac{3}{2}(J_1+J_2+J_3)
\end{bmatrix} .
\end{equation}
The AFM ground state is hence
\begin{eqnarray}
\mspace{-48mu} | AFM_4' \rangle & \! = & \! -\sin \frac{\beta}{2} | S=0, 1 \rangle + \cos \frac{\beta}{2} | S=0, 2 \rangle \\
& \! = & \! \left( \frac{1}{\sqrt{12}} \cos \frac{\beta}{2} - \frac{1}{2} \sin \frac{\beta}{2} \right) | \uparrow, \uparrow, \downarrow, \downarrow \rangle - \frac{1}{\sqrt{3}} \cos \frac{\beta}{2} | \uparrow, \downarrow, \uparrow, \downarrow \rangle + \left( \frac{1}{\sqrt{12}} \cos \frac{\beta}{2} + \frac{1}{2} \sin \frac{\beta}{2} \right) | \downarrow, \uparrow, \uparrow, \downarrow \rangle + \dotsb \label{eq-AFM4}
\end{eqnarray}
with the mixing angle $\beta = \arctan \left( \sqrt{3} \frac{J_1-J_2+J_3}{J_1+J_2+J_3} \right)$ and the energy $E_{AFM'}^{(4)} = E_F^{(4)} - (J_1+J_2+J_3) \left( 1 + \frac{1}{2 \cos \beta} \right)$. Exactly on resonance, the probability for the tunneling of two spin-down particles is therefore given by
\begin{equation}
P_\downarrow(-1/g_\text{1D}=0) = |\langle \uparrow, \uparrow , \downarrow , \downarrow | AFM_4' \rangle|^2 = \left( \frac{1}{\sqrt{12}} \cos \frac{\beta}{2} - \frac{1}{2} \sin \frac{\beta}{2} \right)^2 .
\end{equation}
We obtain for the experimental parameters at the CIR (see Table~\ref{fig-spin_down_table}(c)) an angle of $\beta \approx 21^\circ$ and hence $P_\downarrow(-1/g_\text{1D}=0) \approx 4\%$. Away from resonance, we have to consider the sequential tunneling of two particles from the initial $| AFM_4' \rangle$ state. In the following, we discuss only the case that a spin-down particle has tunnelled in the first tunneling process (the spin-up case is analogous). Therefore, we have to consider the tunneling into the final states $| FM_3' \rangle = | FM_3 \rangle | \downarrow \rangle$, $| IM_3''' \rangle = | IM_3' \rangle | \downarrow \rangle$, and $| AFM_3''' \rangle = | AFM_3' \rangle | \downarrow \rangle$, where $| FM_3 \rangle$ is given by Eq.~(\ref{eq-FM}), $| IM_3' \rangle$ is given by Eq.~(\ref{eq-IM}), and $| AFM_3' \rangle$ is given by Eq.~(\ref{eq-AFM}). It follows from Eqs.~(\ref{eq-FM}) and~(\ref{eq-AFM4}) that the spin overlap of $| AFM_4' \rangle$ with $| FM_3' \rangle$ is exactly zero independent of the tilting of the trap. Moreover, the squared spin overlap with the IM state,
\begin{equation}
|\langle IM_3''' | AFM_4' \rangle |^2 = \frac{1}{2} \sin^2 \left( \frac{\alpha-\beta}{2} \right) ,
\end{equation}
reaches at most $0.1\%$, while the squared spin overlap with the AFM state,
\begin{equation}
|\langle AFM_3''' | AFM_4' \rangle |^2 = \frac{1}{2} \cos^2 \left( \frac{\alpha-\beta}{2} \right) ,
\end{equation}
is at least $49.9\%$ (the remaining $50\%$ belong to the spin-up case). Therefore, even in the super-Tonks regime, where tunneling into the IM state is energetically favored, the particle tunnels with at most $1\%$ into the IM state due to the small spin overlap. We conclude that the system ends up in the AFM three-particle in-trap state with $\approx 50\%$ after the first tunneling process. The second tunneling process is then the tunneling from the AFM state of a (2,1) system, which has been discussed above. It follows that the probability for spin-down tunneling, $P_\downarrow = P_{| AFM_4' \rangle, | \uparrow, \uparrow \rangle}$, is for the (2,2) system half as large as for the (2,1) system in the whole interaction regime. This is indeed the case, as can be seen in Fig.~3(c) of the main text.

\section{X. \quad	Exact diagonalization}

The occupation-number distributions of Fig.~4 of the main text have been calculated by means of an exact diagonalization of the full continuous model. The second-quantized Hamiltonian of the full model reads
\begin{equation}
\hat H = \sum_{\sigma = \downarrow, \uparrow} \int dz \hat \Psi_\sigma^\dagger(z) \left( -\frac{\hbar^2}{2m} \frac{d^2}{dz^2} + \frac{1}{2} m \omega_{||}^2 z^2 \right) \hat \Psi_\sigma(z) + g_\text{1D} \int dz dz' \hat \Psi_\uparrow^\dagger(z) \hat \Psi_\downarrow^\dagger(z') \delta(z-z') \hat \Psi_\downarrow(z') \hat \Psi_\uparrow(z)
\end{equation}
with the fermionic field operators $\hat \Psi_\sigma(z)$ and the angular frequency $\omega_{||}$ of the harmonic oscillator at the local minimum. We expand the field operators in the energy eigenbasis of the harmonic oscillator, $\hat \Psi_\sigma(z) = \sum_i \phi_i(z) \hat a_{i,\sigma}$. Here, $i$ refers to the $i$th energy level, $\phi_i(z)$ is the $i$th energy eigenstate of the harmonic oscillator, and $\hat a_{i,\sigma}$ annihilates a fermion in energy eigenstate $i$ and spin state $\sigma$. The Hamiltonian is then numerically calculated and diagonalized in the many-body energy eigenbasis of the noninteracting problem. Here, the finite size of the computer memory restricts the calculation to all noninteracting eigenstates below a certain cutoff energy. This allows for an accurate numerical calculation of the low-energy eigenstates of the interacting few-body system (for not too many particles and not too strong interactions, $|g_\text{1D}| \leq 20$). The mean occupancy of single-particle level ${(i,\sigma)}$ is the expectation value $\langle \hat a_{i,\sigma}^\dagger \hat a_{i,\sigma} \rangle$ of the corresponding eigenstate of the interacting system.

\end{appendix}

\end{document}